\RequirePackage{lineno} 
\documentclass[12pt]{article}
\usepackage{graphicx,psfrag,epsf,subfigure,adjustbox}
\usepackage{enumerate}
\usepackage[english]{babel}
\usepackage{amssymb,amsmath,amsthm,amsfonts}

\usepackage{mathtools} 

\usepackage{bbm}
\usepackage{color}
\usepackage{lineno}
\usepackage{dsfont}
\usepackage{graphicx}
\usepackage{url}
\usepackage[section]{placeins}
\usepackage{hyperref}
\usepackage{enumitem}
\usepackage{booktabs}
\usepackage{bm}
\usepackage{breqn}
\usepackage[font=small]{caption}

\newcommand{\data}{\bm{x}}
\newcommand{\dataObs}{\bm{x^0}}
\newcommand{\datasim}{\bm{x^{\mbox{sim}}}}
\newcommand{\distance}{d}
\newcommand{\Model}{\mathcal{M}}
\DeclareMathOperator*{\argmin}{arg\,min}
\newcommand{\parameter}{\bm{\theta}}
\newcommand{\parametertheo}{\bm{\phi}}
\newcommand{\estparametertheo}{\bm{\hat{\phi}}}
\newcommand{\prior}{\pi}
\newcommand{\estparameter}{\hat{\bm{\theta}}}
\newcommand{\network}{\mathcal{N}}

\newcommand{\infectednodes}{\mathbb{N}_{\mathbb{I}}}

\newcommand{\exposednodes}{\mathbb{N}_{\mathbb{E}}}
\newcommand{\seednode}{n_{\tiny{\textrm{SN}}}}
\newcommand{\seednodea}{n_{\tiny{\textrm{SN1}}}}
\newcommand{\seednodeb}{n_{\tiny{\textrm{SN2}}}}
\newcommand{\estseednode}{\hat{n}_{\tiny{\textrm{SN}}}}
\newcommand{\ratediffusion}{\theta}
\newcommand{\estratediffusion}{\hat{\theta}}
\newcommand{\rateexposure}{\beta}

\newcommand{\threshold}{\gamma}

\newcommand{\obsstarttime}{t_0}
\newcommand{\obsendtime}{T}
\newcommand{\summaryCC}{\mathbb{L}}
\newcommand{\lossfunc}{\mathcal{L}}
\newcommand{\summaryfunc}{\mathcal{F}}
\newcommand{\distdata}{d}

\newcommand{\shortgraphdist}{d_{\mathbb{N}}}
\newcommand{\eucldist}{d_{E}}
\newcommand{\er}{Erd\H{o}s-R\'{e}nyi }
\newcommand{\ba}{Barab\'{a}si-Albert }
\newcommand{\cardinality}[1]{\left\vert{#1}\right\vert}

\newcommand{\blind}{0}
\begin{document}

\if0\blind
{

\title{\bf Bayesian Inference of Spreading Processes on Networks \thanks{The networks and the codes used to simulate the spreading processes and to infer the spreading process parameters and seed-node from the simulated epidemics can be downloaded from: \url{https://github.com/eth-cscs/abcpy-models/tree/master/Network/SpreadingProcessNetwork.}}} 

\author{Ritabrata Dutta$^1$\thanks{Corresponding author:duttar@usi.ch}, Antonietta Mira$^{1,2}$, Jukka-Pekka Onnela$^3$\\ {\em \small $^1$Institute of Computational Science, Universit\`a della Svizzera italiana, Switzerland}\\ 
{\em \small$^2$Department of Science and High Technology, Universit\`a degli Studi dell'Insubria, Italy}\\  
{\em \small$^3$Department of Biostatistics, Harvard University, USA}
  } \maketitle } \fi

\if1\blind
{
  \bigskip
  \bigskip
  \bigskip
  \begin{center}
    {\LARGE\bf Title}
\end{center}
  \medskip
} \fi

\bigskip
Infectious diseases are studied to understand their spreading mechanisms, 
to evaluate control strategies and to predict the risk and course of future outbreaks. Because people only interact with a small number of individuals, and because the structure of these interactions matters for spreading processes, the pairwise relationships between individuals in a population can be usefully represented by a network. Although the underlying processes of transmission are different, the network approach can be used to study the spread of pathogens in a contact network or the spread of rumors in an online social network. We study simulated simple and complex epidemics on synthetic networks and on two empirical networks, a social / contact network in an Indian village and an online social network in the U.S. Our goal is to learn simultaneously about the spreading process parameters and the source node (first infected node) of the epidemic, given a fixed and known network structure, and observations about state of nodes at several points in time. Our inference scheme is based on approximate Bayesian computation (ABC), an inference technique for complex models with likelihood functions that are either expensive to evaluate or analytically intractable. ABC enables us to adopt a Bayesian approach to the problem despite the posterior distribution being very complex. Our method is agnostic about the topology of the network and the nature of the spreading process. 
It generally performs well and, somewhat counter-intuitively, the inference problem appears to be easier on more heterogeneous network topologies, which enhances its future applicability to real-world settings where few networks have homogeneous topologies.

\vspace{1ex}\noindent
\vfill

\section{Introduction}
\label{sec:intro}
Human susceptibility to epidemics of misinformation and disease has grown manyfold as the world we inhabit keeps getting smaller due to increased access to online information and soaring global mobility \cite{RePEc:ess:wpaper:id:10737}. Social media platforms have changed the way we consume information \cite{Schmidt21032017}, and more and more people find their news through social media \cite{newman2015reuters}. Following the 2016 presidential election in the United States, there have been investigations into the spread of false stories, or ``fake news'' on social media, and based on web browsing data, archives of fact-checking websites, and results from an online survey, a recent study found that social media were an important source of election news \cite{allcott2017social}. 

While ascertainment of social network structures is generally difficult using traditional survey-based approaches, such as name generators, which are survey questions designed  to solicit information about friends and acquaintances of a subject, online platforms readily capture the structure of large-scale social networks, therefore making them 
well suited to study spread of information whether accurate or not. Further, although the transmission mechanisms are very different, the spread of information in online systems has many similarities to the spread of infectious diseases among hosts in a population. From a mathematical and statistical point of view, one can therefore investigate the spread of pathogens and the spread of information in the same framework as long as the network structure accurately captures the transmission pathways and the spreading process is parametrized appropriately.

In this paper, we consider a simple susceptible-infected (SI) process and a more complex spreading process on a fixed and known network structure. This spreading process may be conceptualized as propagating either a pathogen or a piece of information. We focus on addressing two distinct questions that are relevant in both settings: (1) How to infer the unknown parameters associated with the spreading process? and (2) How to identify the source node that most likely was responsible for introducing the pathogen or information into the network? To address these questions, we assume that the system evolves in discrete time steps, and we observe the state of all nodes
at each time steps in the course of the epidemic from time $\obsstarttime$ (first observation) to time $\obsendtime$ (last observation). We further assume that the spreading process is parametrized by $\ratediffusion$ and the epidemic starts from a single node, which we call the seed node, denoted 
by $\seednode$. Our goal is to perform inference jointly on $\ratediffusion$ and $\seednode$ given the observed data. Because the spreading process operates on a network, the likelihood function for the data is not analytically available, and we therefore use approximate Bayesian computation (ABC) to simulate samples from the approximate posterior distribution of $(\ratediffusion, \seednode)$ given the observed data. 

Bayesian inference of spreading process parameters is a well developed research area \cite{streftaris2002statistical, o2002tutorial, demiris2005bayesiana, demiris2005bayesianb, okamura2007statistical}, where ABC has been used to infer parameters of spreading processes in recent past \cite{toni2009approximate, neal2012efficient, brooks2014dynamic, kypraios2017tutorial}. To our knowledge, for a spreading process propagated on a network, Bayesian inference has only been used to infer parameters of a susceptible-infected-recovered (SIR) epidemic model on a Bernoulli random graph \cite{britton2002bayesian, neal2005case, kypraios2017tutorial}. 
This is the first attempt to extend the Bayesian framework to different spreading processes on different network topologies, whereas we acknowledge that inference of parameters for SIR process is harder than for SI process (being a special case of SIR) and equally, if not less difficult than for complex contagion process, considered in this manuscript.
 Separate from this line of research, 
the estimation of the seed-node, learning which nodes were infected at a fixed time point, has been studied in  
\cite{lappas2010finding,shah2011rumors, prakash2012spotting}. Here, we give a framework where we can simultaneously estimate the spreading process parameters and the seed-node, which can be applied to any contagion process and any general network topology. To illustrate the generality of our method in Section~\ref{sec:result}, we compare the performance of suggested Bayesian parameter estimates with the Netsleuth algorithm \cite{prakash2014efficiently} for seed-node detection.

We study this problem on synthetic network topologies generated by the \ba (BA) and \er (ER) models. Finally, to investigate the performance of our inference scheme on empirical networks, we consider the simulated spread of a pathogen among the inhabitants of a rural Indian village and the simulated spread of information among a group of Facebook users. Because of the differences in spreading dynamics, we use two similar but distinct spreading processes, simple and complex. As the results demonstrate, our general inference framework performs reasonably well across a range of network topologies, whether synthetic or empirical, and for both types of spreading processes (simple and complex).

\section{Spreading processes}
\label{sec:networkmodel}

\subsection{Simple contagion}
For modeling the spread of infections on a human contact network, we consider a simple spreading process, i.e., the standard susceptible-infected (SI) process on a fixed network. In this model, there are only two states, susceptible and infected, and this process is suitable for modeling the spread of pathogens in contact networks because a single successful exposure can be sufficient for transmission. In our specification of the process with unit infectivity \cite{zhou2006behaviors,staples2016leveraging}, at each time step, each infected node chooses one of its neighbors with equal probability regardless of their status (susceptible or infected), and if the chosen node is susceptible, it is infected with probability $\ratediffusion$. We denote this model by $\Model_S$ and parametrize it in terms of the spreading rate $\ratediffusion$ and of the seed node $\seednode$. 
For given values of these two parameters, $\seednode = \seednode^* \mbox{ and } \ratediffusion = \ratediffusion^*$, we can forward simulate the evolving epidemic over time using the $\Model_S$ model as
\begin{eqnarray}
\label{eq:simulator_SC}
\Model_S [\seednode^*, \ratediffusion^*]  \rightarrow \lbrace \infectednodes(t), t=0, 1, \ldots, \obsendtime \rbrace,
\end{eqnarray}
where $\infectednodes(t)$ is a list of infected nodes at time $t$. 

\subsection{Complex contagion}
Our complex spreading process incorporates the idea that 
social, unlike biological contagion, may require multiple exposures, possibly from multiple hosts, to render a susceptible person infected, where, in this context, an infected person refers to someone who has adopted the circulating piece of information and can spread it among his or her contacts. Non-probabilistic threshold models for complex contagion have a long history \cite{GranovetterThreshold_1978}, 
and they have been applied to datasets capturing the spread of moods, ideas and news on social networks \cite{hill2015spreading, Fink_AAAI_2016, eyre2017spreading} and their theoretical properties are 
well known \cite{centola2007complex, centola2010spread, barash2012critical}. A stochastic version of this threshold-based complex contagion process was recently developed in \cite{Fink_AAAI_2016}, and it was shown to perform well in modeling a large Twitter dataset. This model has three states: susceptible (S), exposed (E), and infected (I), where a susceptible person (S) has not been exposed to the piece of information, the exposed person (E) has been exposed to the information but is not yet infectious, and finally the infected person (I) is able to spread the information further. Hence, the complex spreading process consists of two sub-processes: transition of a susceptible person to an exposed one and transition of an exposed person to an infected one.

As our goal here is not to find the best model suitable for a specific dataset, but rather to illustrate the properties of our inferential framework, we introduce some modifications to the model of \cite{Fink_AAAI_2016}. Our model corresponds to the model of  \cite{Fink_AAAI_2016} as far as becoming infected (E to I transition) is concerned, but to model the exposure (S to E transition), we use the SI process with unit infectivity. 
The process of how a person becomes exposed is likely to vary across settings, but we absorb these multiple factors in a single probability, over a time window (e.g., a single day), and we use this as the basis of our exposure mechanism. More specifically, we use a Bernoulli random variable parametrized by the \textit{rate of exposure} $\rateexposure$ to model the process by which a susceptible neighbor of an infected node becomes exposed. 

In this complex spreading process, the transition of an exposed person to an infected one depends both on how many times she has been exposed to the information and how many of her friends (neighboring nodes) are infected.
To keep track of exposures, we define a collection of exposure summaries $\summaryCC(t)$ for all nodes, where the exposure summary for node $i$ at time $t$, denoted by $\summaryCC_i(t)$, has length equal to the number of infected neighbors of node $i$. The first element of $\summaryCC_i(t)$,
that we denote by  $\summaryCC_{i,1}(t)$,
 counts the number of exposures node $i$ experienced when it had a single infected neighbor, and, in general, the $k^{th}$ element, denoted by  $\summaryCC_{i,k}(t)$, 
counts the number of exposures the node experienced when it had $k$ infected neighbors. The length of these summary statistics generally increases with time. For example, let us consider the following summary list for node $i$ at time $t=4$:
\begin{dmath*}
\summaryCC_i(4) = (3,5). 
\end{dmath*}
Here the first element,  
$\summaryCC_{i,1}(4) = 3$, 
indicates that the node had 3 exposures when it had a single infected neighbor, and the second element,  
$\summaryCC_{i,2}(4)=5$, 
indicates that the node had 5 exposures when it had 2 infected neighbors. 
Let $|\summaryCC_i(t)|$ denote the length of the 
vector $\summaryCC_i(t)$ which corresponds to the number of infected neighbors node $i$ has by time $t$.
In the example above $|\summaryCC_i(4)| = 2$ meaning that, at time $t=4$ node $i$ had 2 infected neighbors.
The probability for node $i$ to become infected precisely at the very last exposure in the vector of exposures captured in $\summaryCC_i(t)$ is given by
\begin{dmath*}
\label{eg:adapt_prob}
\left(\prod_{k=1}^{|\summaryCC_i(t)|-1}(1-p_k)^{\summaryCC_{i,k}(t)}\right)(1-p_{|\summaryCC_i(t)|})^{\summaryCC_{i,|\summaryCC_i(t)|}(t)-1}p_{|\summaryCC_i(t)|}
\end{dmath*}
where $p_k$ is the probability of infection when a node has $k$ infected neighbors.
To approximate the threshold-like behavior of the non-probabilistic complex contagion model
of \cite{barash2012critical},
$p_k$ is modeled as a modified logistic sigmoid function:
\begin{dmath*}
\label{eq:adapt_lik}
p_k = \epsilon_{\textrm{low}} + \frac{\epsilon_{\textrm{high}}-\epsilon_{\textrm{low}}}{1+\exp\left({-g(k- \threshold F_i)}\right)}
\end{dmath*} 
where $k>0$ and $F_i$ is the degree of node $i$, i.e., the number of neighbors of $i$. 
We note that $p_k$ is an increasing function of $k$. The minimum and maximum values of $p_k$ $(\epsilon_{\textrm{high}}$ and $\epsilon_{\textrm{low}})$, for all $k\geq 1$ are fixed to $0.001$ and $0.25$, respectively, and the shape parameter is $g=1$. The \textit{threshold} parameter $\threshold$ shifts the location of the threshold in the sigmoid function, approximating the effect of the relative threshold in the non-probabilistic complex contagion model. In a non-probabilistic complex contagion model, an exposed person could be infected only after a fraction $\threshold$ of the neighboring nodes has already been infected. Similarly, here the probability of infection $p_k$ is significantly higher when $k>\threshold F_i$ i.e. if 
a fraction greater than $\threshold$ of neighboring nodes are infected. 

To initiate the spreading process from a given seed node $\seednode=\seednode^*$, we infect a proportion of its neighbors (following \cite{centola2007complex} we take this proportion to be $\threshold$) and these infected nodes constitute the first wave of the infection; the contagion is thereafter propagated throughout the network until all nodes are infected. (More precisely, when $\seednode=\seednode^*$, we first round the product $\threshold F_{\seednode^*}$ to the nearest integer, choose this number of nodes at random from among the neighbors of node $\seednode^*$, and then deterministically infect all of these nodes.) The complex contagion model, denoted by $\Model_C$, is parametrized in terms of the \textit{rate of exposure} $\rateexposure$ and \textit{threshold} $\gamma$, collectively denoted by $\parameter = (\rateexposure, \threshold)$. Given the model $\Model_C$ and values of $\seednode=\seednode^*$ and $\parameter=\parameter^*$, we can forward simulate the spreading process as 
\begin{eqnarray}
\label{eq:simulator_CC}
\Model_C [\seednode^*, \parameter^*]  \rightarrow \left\lbrace \left(\infectednodes(t),\exposednodes(t),\summaryCC(t)\right), \ t=0,1, \ldots, \obsendtime \right\rbrace,
\end{eqnarray}
where the model generates the output lists $\infectednodes(t),\ \exposednodes(t) \ \mbox{and} \ \ \summaryCC(t)$, which correspond to the infected and exposed nodes in the network, plus a collection of exposure summaries for all nodes $i\in \exposednodes(t)$ at time $t$. 

\section{Data generation}
\subsection{Simulating observed data}
We use the term \emph{observed data}, denoted by $\dataObs$, to refer to a dataset generated by some real-world process, and our goal is to learn about parameter values characterizing this process. 
In this paper, 
for illustrative purposes, we generate the observed datasets using the models described above. 
If we assume that the first observation of the process occurs at time $\obsstarttime$ and the last observation at time $\obsendtime$, then the observed dataset for the simple contagion model is $\dataObs \equiv \lbrace \infectednodes^0(t): t =\obsstarttime, \obsstarttime+1, \ldots, \obsendtime \rbrace$ and for the complex contagion model $\dataObs \equiv \lbrace (\infectednodes^0(t), \exposednodes^0(t),\summaryCC^0(t)): t =\obsstarttime, \obsstarttime+1, \ldots, \obsendtime \rbrace$. Given $\dataObs$ and the fixed structure of the underlying network, the goal is to estimate the model parameters and the identity of the source-node. To this aim, we develop below an approximate Bayesian inference scheme that allows us to quantify the uncertainty in the inferred model parameters, uncertainty which is inherent to the inferential process given the stochastic nature of the models described in Equations~\ref{eq:simulator_SC} \& \ref{eq:simulator_CC}.

\subsection{Synthetic networks}
The topology of social and contact networks can be significantly different and epidemic spreading processes are usually sensitive to the topology of the network. To study our inference scheme on different types of networks, we generated synthetic networks using the \ba (BA) \cite{barabasi1999emergence} and the \er (ER) \cite{erdos1959random} models, where in each case the networks 
have 100 nodes. 
The difficulty of inference is affected by the number of infected nodes at different times, which depends on several factors, such as network structure, network size, spreading process (including its parameters). To make the problem more computationally tractable, we study the inference scheme in details on  synthetic networks of relatively small size (100 nodes) and larger empirical networks. The BA model is characterized by two parameters, one of which is network size, and we set the other parameter $m=4$, which specifies that each new node gives rise to $m=4$ edges in the network. The network is grown from a small seed network by the addition of nodes, one at a time, and the edges are
 inserted using proportional preferential attachment, meaning they are 
connected to existing nodes with a probability that is proportional to the degree of already existing nodes \cite{easley2010networks}.
The ER model is likewise characterized by two parameters, one of which is also network size, and we set the other parameter $p=0.05$, which specifies the probability that any  two nodes are connected with an edge, i.e., each dyad (pair of nodes) is associated with an IID Bernoulli random  variable. We considered both simple and complex contagion processes on networks generated from these models with the following details:
\subsubsection*{Simple contagion} A simple contagion epidemic is simulated on the BA and ER networks starting from a seed node $\seednode^0$ that is chosen uniformly at random over all network nodes and the common diffusion rate is set to $\ratediffusion^0 = 0.3$. The observed dataset $\dataObs$ is constructed from the simulated dataset by setting $\obsstarttime = 20$ and  $\obsendtime = 70$.
\subsubsection*{Complex contagion} For the simulation of complex contagion epidemics, the values of the parameters 
are fixed at $\parameter^0 \equiv (\rateexposure^0, \threshold^0) = (0.7, 0.3)$. For each simulation round, we use the same network realization that was generated for the simple contagion and we also start the epidemic from the same seed node used in the simple contagion. The observed dataset $\dataObs$ is constructed by setting $\obsstarttime = 20$ and  $\obsendtime = 120$.

\subsection{Empirical networks}
We also consider two empirical networks, a human social / contact network constructed among 354 inhabitants of a village of South Indian state of Karnataka \cite{banerjee2013diffusion} and a social network among 4039 Facebook users \cite{leskovec2012learning}.

\subsubsection*{Simple contagion of disease in an Indian village} We consider a simple contagion model to simulate an epidemic of a disease in the Indian village contact network. The network has 354 nodes and 1541 edges, representing 354 villagers and reported contacts and social relationships among them. The epidemic is simulated using $\ratediffusion^0 = 0.3$, $\seednode^0=70$, and the observed dataset $\dataObs$ is the infected nodes $\infectednodes(t)$ for $t = \obsstarttime, \obsstarttime+1, \ldots, \obsendtime$ with $\obsstarttime=20$ and $\obsendtime=70$. 
\subsubsection*{Complex contagion of information on Facebook} To simulate an epidemic of information on a Facebook network, we use the complex contagion model. The network has 4039 nodes and 88234 edges; other topological features of the network can be found in \cite{leskovec2012learning}. For the complex contagion model, we fix the parameter values as $\parameter^0\equiv(\rateexposure^0,\threshold^0)=(0.7,0.3)$, $\seednode^0=2000$, and the observed dataset $\dataObs$ is the infected nodes $\infectednodes(t)$, exposed nodes $\exposednodes(t)$, and a collection of exposure summaries for all exposed nodes $\summaryCC{(t)}$  at time $t = \obsstarttime, \obsstarttime+1,\ldots, \obsendtime$, when $\obsstarttime = 20$ and $\obsendtime = 120$. 

For the sake of completeness, we also consider the complex and simple contagion process on Indian village contact network and Facebook network respectively starting at $\seednode^0=70$ and $\seednode^0=2000$. The epidemics are simulated in these cases using the same values for the other parameters used to simulate the complex and simple contagion processes correspondingly on Facebook and Indian village contact network above.

\section{Inference framework}
\label{sec:if}

\subsection{Bayesian inference}
\label{sec:BI}
We can quantify the  uncertainty of the inferred parameter $\parametertheo = (\parameter,\seednode)$ by their posterior density $p(\parametertheo|\data)$ given the observed dataset $\data = \dataObs$. The posterior density can be written by Bayes' theorem as,
\begin{eqnarray}
p(\parametertheo|\data) = \frac{\prior(\parametertheo)p(\data|\parametertheo)}{m(\data)},
\end{eqnarray}
where $\prior(\parametertheo)$, $p(\data|\parametertheo)$ and $m(\data) = \int\prior(\parametertheo)p(\data|\parametertheo)d\parametertheo$ are, correspondingly, the prior density on the parameter $\parametertheo$, the likelihood function, and the marginal likelihood. The prior density $\prior(\parametertheo)$ enables a way to leverage the learning of parameters from prior knowledge, which is expected to be available especially for infectious diseases. If the likelihood function can be evaluated, at least up to a normalizing constant, then the posterior density can be approximated by drawing a sample of parameter values from it using (Markov chain) Monte Carlo sampling schemes \cite{Robert2005}. The likelihood function induced by our model of network epidemics is too demanding computationally to evaluate due to the complex stochastic nature of the models. 

In this setting, approximate Bayesian computation (ABC) \cite{lintusaari2017fundamentals} offers a way to sample from the approximate posterior density and opens up the possibility of sound statistical inference on the parameter $\parametertheo$.

\subsection{Approximate Bayesian computation (ABC)}
\label{sec:Inference_method}
Models that are easy to forward simulate, given values of the parameters, are called simulator-based models in the ABC literature\footnote{In this manuscript, we will use the term simulator-based model to refer to a model that enables direct simulation of model outcomes using a set of stochastic rules. This term is well established within the ABC literature, but we point out that these types of models are sometimes called mechanistic models or agent based models in different fields of science.} and they are used in a wide range of scientific disciplines to describe and understand different aspects of nature ranging from dynamics of sub-atomic particles \cite{Martinez_2016} to evolution of human societies \cite{Turchin_2013} and formation of universes \cite{Schaye_2015}.
In the fundamental rejection ABC sampling scheme, we simulate a synthetic dataset $\datasim$ from the simulator-based model $\Model(\parametertheo)$ for a fixed parameter value of $\parametertheo$ and measure the closeness between $\datasim$ and $\dataObs$ using a pre-defined discrepancy measure $\distance(\datasim,\dataObs)$. Based on this discrepancy measure, ABC accepts the parameter value $\parametertheo$ when $\distance(\datasim,\dataObs)$ is less than a pre-specified threshold value $\epsilon$. 

The intractable likelihood $\mathcal{L}(\parametertheo)$ is approximated by $\mathcal{L}_{\distance,\epsilon}(\parametertheo)$ for some $\epsilon>0$, where 
\begin{eqnarray}
\label{eq:approx_lik}
\mathcal{L}_{\distance,\epsilon}(\parametertheo) \propto P(\distance(\datasim,\dataObs)<\epsilon)
\end{eqnarray}
and, as a consequence, the sampled parameters follow the posterior distribution of $\parametertheo$ conditional on $\distance(\datasim,\dataObs)<\epsilon$:
\begin{eqnarray}
\label{eq:approx_post}
p_{\distance,\epsilon}(\parametertheo|\dataObs) \propto P(\distance(\datasim,\dataObs)<\epsilon)\prior(\parametertheo).
\end{eqnarray}
For a better approximation of the likelihood function, computationally efficient sequential ABC algorithms \cite{Marin_2012} 
decrease the value of the threshold $\epsilon$ adaptively while exploring the parameter space.
Here, we use simulated annealing approximate Bayesian computation (SABC) \cite{Albert_2015} implemented in our ABCpy Python package \cite{Dutta_2017_PASC} for optimal utilization of an HPC environment. 
Using SABC, we can draw samples from $p_{\distance,\epsilon}(\parametertheo|\dataObs)$ [Equation~\ref{eq:approx_post}], for a very small value of $\epsilon$ such that the distribution numerically converges to the posterior distribution $p(\parametertheo|\dataObs)$.
All the tuning parameters for the SABC algorithm are fixed at the default values in the ABCpy package with the exception of the number of steps and the acceptance rate cutoff, which were set to 200 and $0.0001$, respectively. We define below the summary statistics extracted from the dataset, the discrepancy measure used, the prior distribution of parameters, and the perturbation kernel to explore the parameter space in the SABC algorithm. 

Generally for ABC, the crucial aspect for a good approximation to the likelihood function is the choice of the summary statistics,
as we define the discrepancy measure between $\datasim$ and $\dataObs$ through a distance between the extracted summary statistics from $\datasim$ and $\dataObs$. 
When choosing the summary statistics, one faces a trade-off between minimizing the loss of information on $\parametertheo$ contained in the data and picking low-dimensional summaries to avoid curse of dimensionality \cite{fearnhead2012constructing}.
Here we use intuitive discrepancy measures between interpretable summary statistics (e.g., the Euclidean distance between the proportion of infected nodes at different time-steps for estimation of $\ratediffusion$) additional to domain-driven discrepancy measures (e.g., the shortest path length distance between the subgraphs induced by the infected nodes, at each time step). Subjectivity of these decisions can be removed through automatic summary selection for ABC, described in \cite{fearnhead2012constructing}, \cite{pudlo2015reliable}, \cite{jiang2015learning} and \cite{gutmann2017likelihood}, where an informative linear or non-linear combination of the summaries is chosen. To keep our summary statistics interpretable, we stick to the intuitive and domain-driven summary statistics and discrepancy measures described next.

\subsubsection*{Summary statistics} Given a dataset $\data \equiv \lbrace \infectednodes(t): t = \obsstarttime,\obsstarttime+1, \ldots, \obsendtime \rbrace$ for the simple contagion model or a dataset $\data \equiv \lbrace \infectednodes(t), \exposednodes(t), \summaryCC{(t)} : t = \obsstarttime,\obsstarttime+1, \ldots, \obsendtime \rbrace$ for the complex contagion model, we compute an array of summary statistics, 
\begin{align*}
&\mbox{Simple contagion, }\summaryfunc_{S}: \data \rightarrow (\bm{I})\\
&\mbox{Complex contagion, }\summaryfunc_{C}: \data \rightarrow (\bm{I}, \bm{E}, \bm{fe})
\end{align*}

defined as follows:
\begin{itemize}
\item $\bm{I} = (I_{\obsstarttime},I_{\obsstarttime+1},\ldots,I_{\obsendtime})$: $I_t$ is the subgraph induced by the infected nodes, at each time $t$.
\item $\bm{E} = (E_{\obsstarttime},E_{\obsstarttime+1},\ldots,E_{\obsendtime})$: $E_t$ is the subgraph induced by the exposed nodes, at each time $t$.
\item $\bm{fe} = (fe_{\obsstarttime},fe_{\obsstarttime+1},\ldots,fe_{\obsendtime})$: $fe_t$ is the proportions of the nodes exposed for the first time, at each time $t$.
\end{itemize}

We note that the summary statistics chosen here are high dimensional, but they are also highly dependent on each other. The subgraphs induced by infected or exposed nodes at time $t_i$ contain the subgraphs induced at the previous time step $t_{i-1}$, but the subgraphs at both the time points are needed to capture information about the evolution of these subgraphs. Similarly the vector containing the proportions of first time exposed nodes, at each time step, is also an increasing sequence of real numbers. This high dependence between the summary statistics reduces their effective dimension. This intuitively explains why the inference scheme does not suffer from the curse of dimensionality, despite having a large number of summary statistics. 

\subsubsection*{Discrepancy measure} The discrepancy measure between two datasets $\data^{(1)}$ and $\data^{(2)}$ is constructed by considering the following two distance functions between the summary statistics extracted from them.\\
Simple contagion:
\begin{eqnarray*}
\label{eq:discrep_measure_SC}
\distdata_{S}(\data^{(1)}, \data^{(2)})
:=\distdata\left(\summaryfunc_{S}(\data^{(1)}), \summaryfunc_{S}(\data^{(2)})\right)
= \shortgraphdist\left(\bm{I}^{(1)},\bm{I}^{(2)}\right) 
\end{eqnarray*}
Complex contagion:
\begin{eqnarray*}
\distdata_{C}(\data^{(1)}, \data^{(2)})
:=\distdata\left(\summaryfunc_{C}(\data^{(1)}), \summaryfunc_{C}(\data^{(2)})\right) =
\shortgraphdist\left(\bm{I}^{(1)},\bm{I}^{(2)}\right)+ \shortgraphdist \left( \bm{E}^{(1)},\bm{E}^{(2)}\right)
+ d_E\left(\bm{fe}^{(1)},\bm{fe}^{(2)}\right)
\end{eqnarray*}
Here $\eucldist$ is the Euclidean distance, 
\begin{align*}
    \shortgraphdist \left( \bm{S}^{(1)},\bm{S}^{(2)} \right)=&\eucldist\left(V\left(\bm{S}^{(1)}\right), V\left(\bm{S}^{(2)}\right) \right)\\
& +   \begin{cases}
       \frac{1}{\obsendtime-\obsstarttime}\sum \limits_{t=\obsstarttime}^{\obsendtime} \frac{1}{c(t)}\sum \limits_{i \in S^{(1)}_t \setminus S^{(2)}_t} \sum \limits_{j \in S^{(2)}_t \setminus S^{(1)}_t } \frac{\rho(i,j)}{\rho_{\max}}, & \text{if}\ \bm{S}^{(1)} \not\subset \bm{S}^{(2)} \ \text{or} \ \bm{S}^{(2)}\not\subset \bm{S}^{(1)}\\
1, & \text{otherwise}
    \end{cases}
  \end{align*}
$V\left(\bm{S}\right) = \left[\cardinality{S_{\obsstarttime}}, \cardinality{S_{\obsstarttime+1}},\ldots,\cardinality{S_{\obsendtime}} \right]/\cardinality{\network}$, $\cardinality{\network}$ is the number of nodes on network $\network$,
$c(t)=\max\left(1,\cardinality{S^{(1)}_t \setminus S^{(2)}_t}\right) \times \max\left(1,\cardinality{S^{(2)}_t \setminus S^{(1)}_t}\right)$, $\rho(i,j)$ is the shortest path length between nodes $i$ and $j$ on the network $\network$ and $\rho_{\max}$ is the maximum shortest path length between any two nodes in the network $\network$, i.e., the network diameter. The motivation for these discrepancy measure 
is to simultaneously capture differences in the proportions of the nodes in different epidemic states as well as differences in their locations on the network.

\subsubsection*{Prior} We use independent prior distributions on $\ratediffusion$, $\rateexposure$, $\threshold$ and $\seednode$: 
\begin{itemize}
\item Prior on $\seednode$: Discrete uniform distribution on the infected nodes of the network at $\obsstarttime$.
\item Prior on  $\ratediffusion$, $\rateexposure$ and $\threshold$: Continuous uniform distribution on $[0,1]$.
\end{itemize}

\subsubsection*{Perturbation kernel} A perturbation kernel used to explore the parameter space is defined as a distribution on $\parametertheo$ conditional on the present parameter values $\parametertheo^*$.\\
Simple contagion:
\begin{align*}
K_S \left(\parametertheo|\parametertheo^*\right) = K_{1S}(\ratediffusion|\ratediffusion^*,\hat{\sigma})K_2(\seednode|\seednode^*)
\end{align*}
Complex contagion:
\begin{align*}
K_C \left(\parametertheo|\parametertheo^*\right) = K_{1C}(\parameter|\parameter^*,\hat{\Sigma})K_2(\seednode|\seednode^*)
\end{align*}
Here $K_{1S}$ and $K_{1C}$ are univariate and multivariate Gaussian distributions  with $\hat{\sigma}$ and $\hat{\Sigma}$, respectively being the estimated variance of $\ratediffusion$ and the variance-covariance matrix of  $\parameter$, sampled in the previous updating step of SABC. 
$K_2$ is a discrete distribution on the neighboring nodes of $\seednode^*$ with each node having a probability inversely proportional to its degree, which makes an uniform exploration
of the network nodes possible irrespective of the heterogeneity present in the network.

\subsection{Parameter estimation}
\label{sec:parameter_estimation}
Given an epidemic on a network, our main goal is to estimate $\parametertheo$ given $\dataObs$. In decision theory, the Bayes estimator minimizes posterior expected loss, $E_{p(\parametertheo|\dataObs)}(\lossfunc(\parametertheo,\cdot)|\dataObs)$, with the following loss functions.\\
Simple contagion:
\begin{align*}
\lossfunc(\parametertheo_1,\parametertheo_2) := d_E(\ratediffusion_1,\ratediffusion_2)+\rho(\seednodea,\seednodeb)
\end{align*}
Complex contagion:
\begin{align*}
 \lossfunc(\parametertheo_1,\parametertheo_2) := d_E(\parameter_1,\parameter_2)+\rho(\seednodea,\seednodeb)
\end{align*}
If we have $Z$ IID samples $(\parametertheo_{i})_{i=1}^{Z}$ from the posterior distribution $p(\parametertheo|\dataObs)$, the Bayes estimator can be approximated as
\begin{eqnarray}
\label{eq:Bayes_estimate}
\estparametertheo = \argmin_{\parametertheo} \frac{1}{Z}\sum_{i=1}^Z \lossfunc(\parametertheo_{i},\parametertheo). 
\end{eqnarray}

\section{Results}
\label{sec:result}

\subsection{Inference on synthetic networks}
\label{sec:infer_simulate}

\begin{figure}[htbp]
    \centering
    \mbox{}%
    \adjustbox{valign=T}{\subfigure{a.}}
    \adjincludegraphics[valign=T,scale=0.2]{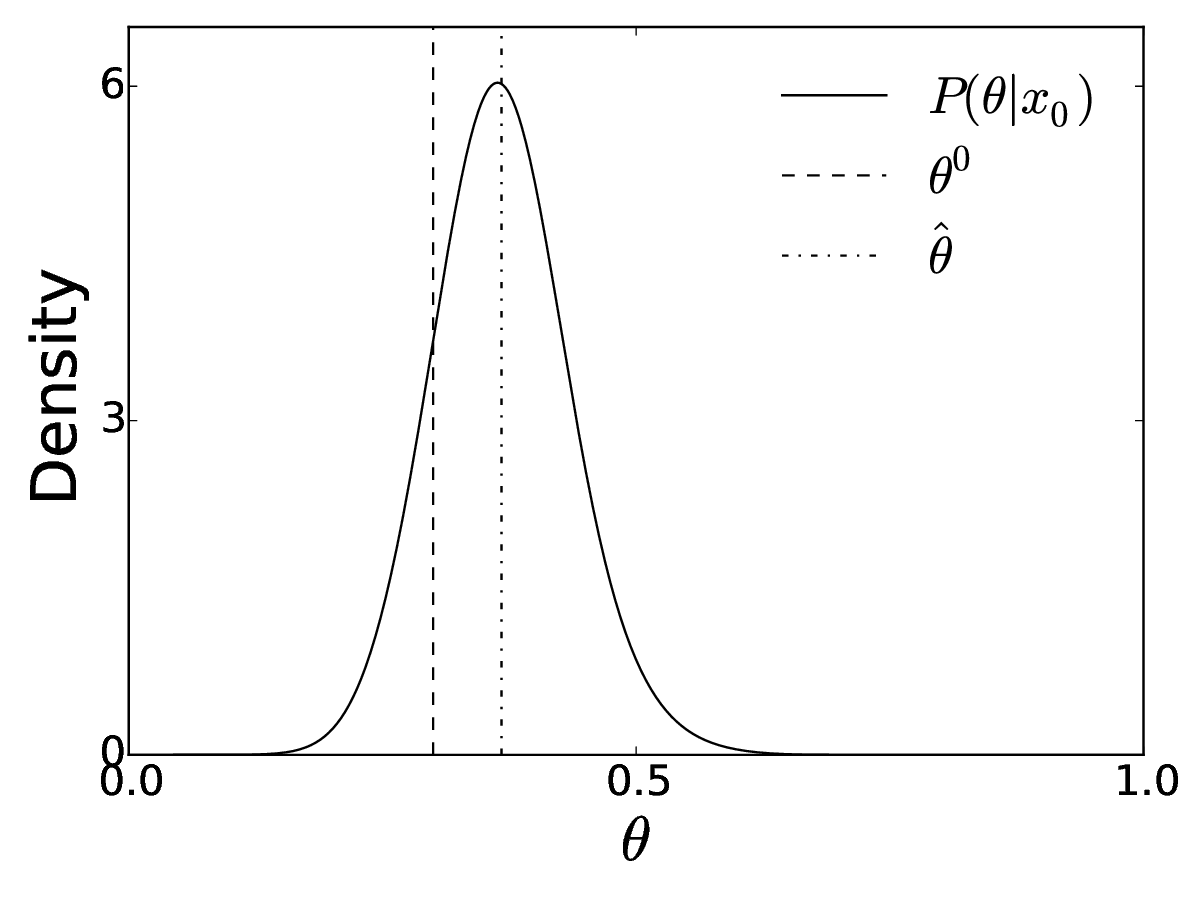}
    \label{fig:ba_SC_theta_posterior}
    \adjustbox{valign=T}{\subfigure{b.}}
    \adjincludegraphics[valign=T,scale=0.2]{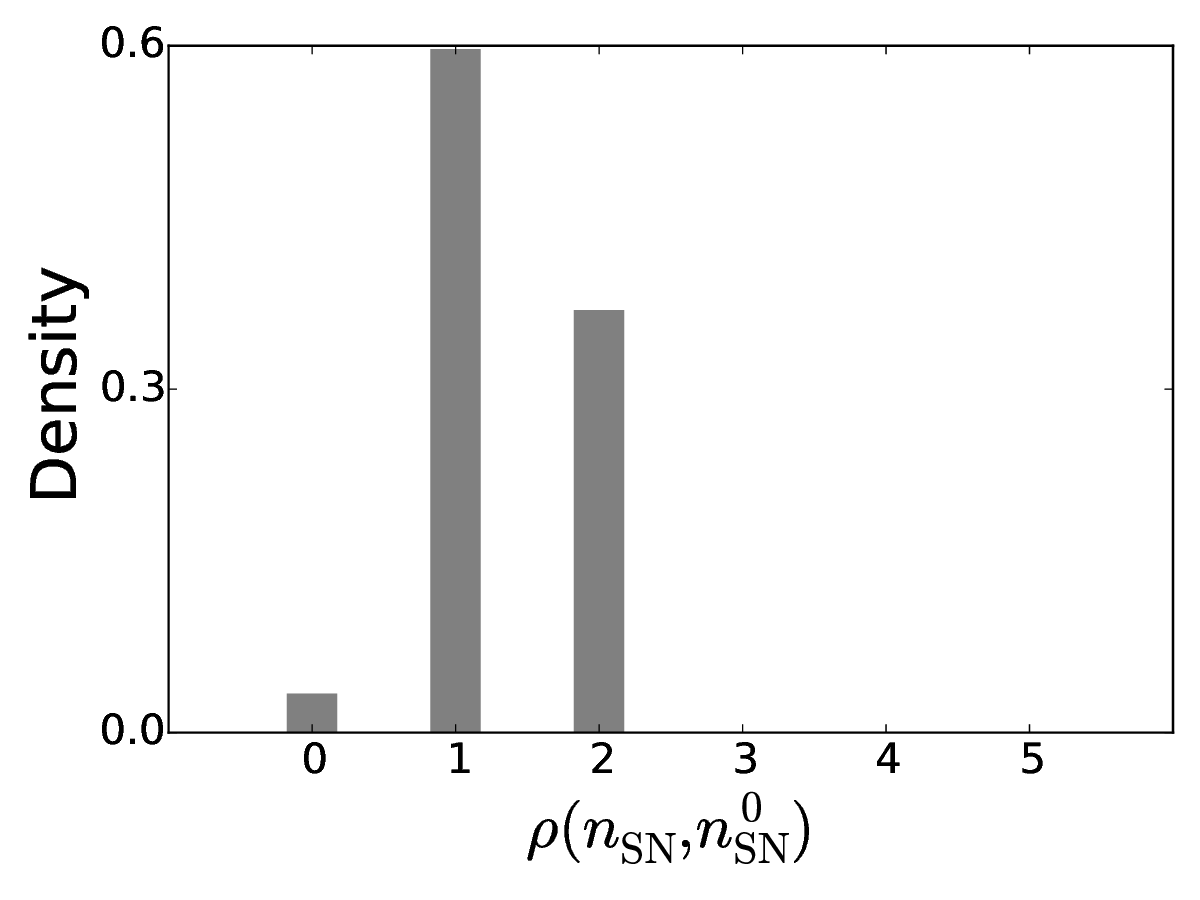}  
    \label{fig:ba_SC_seednode_posterior}
    \adjustbox{valign=T}{\subfigure{c.}}
    \adjincludegraphics[valign=T,scale=0.2]{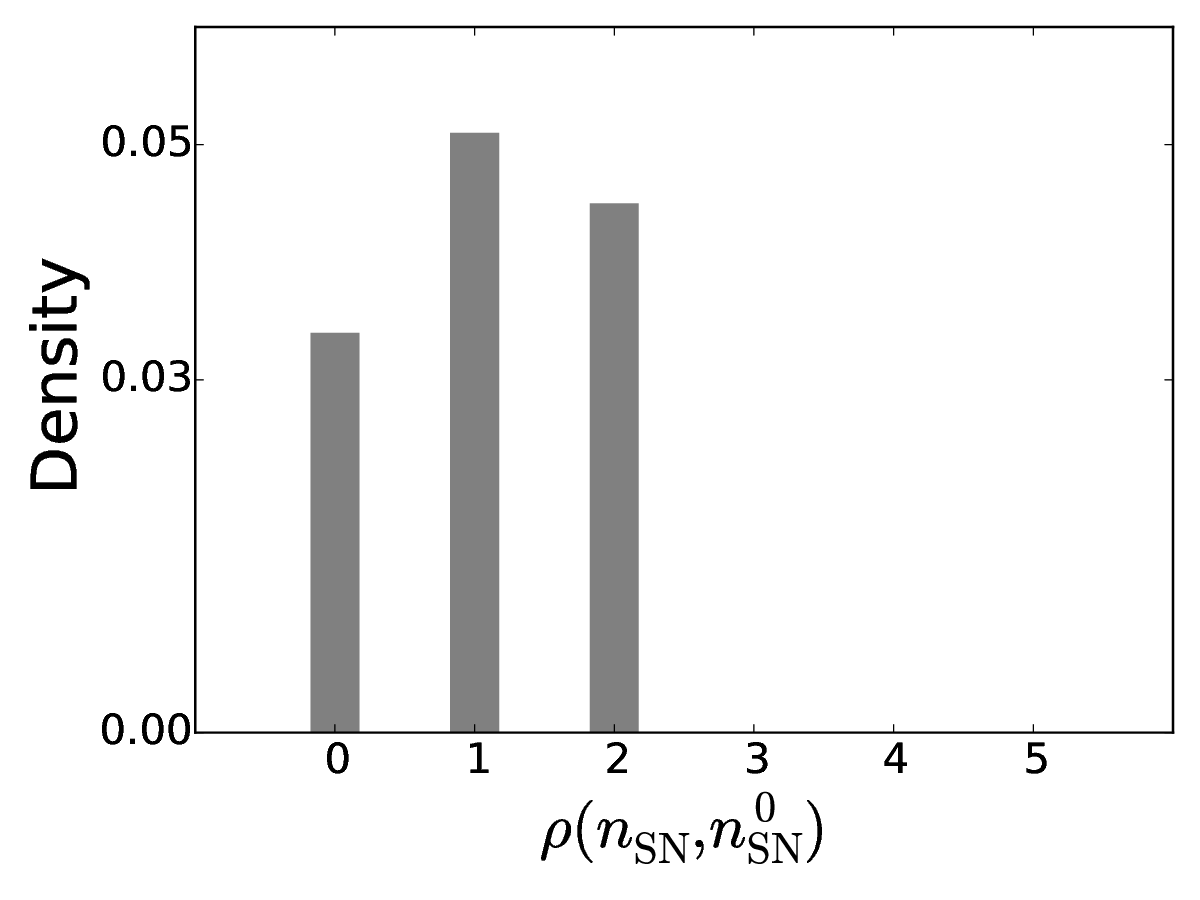}  
    \label{fig:ba_SC_seednode_posterior_max}

    \adjustbox{valign=T}{\subfigure{d.}}
    \adjincludegraphics[valign=T,scale=0.2]{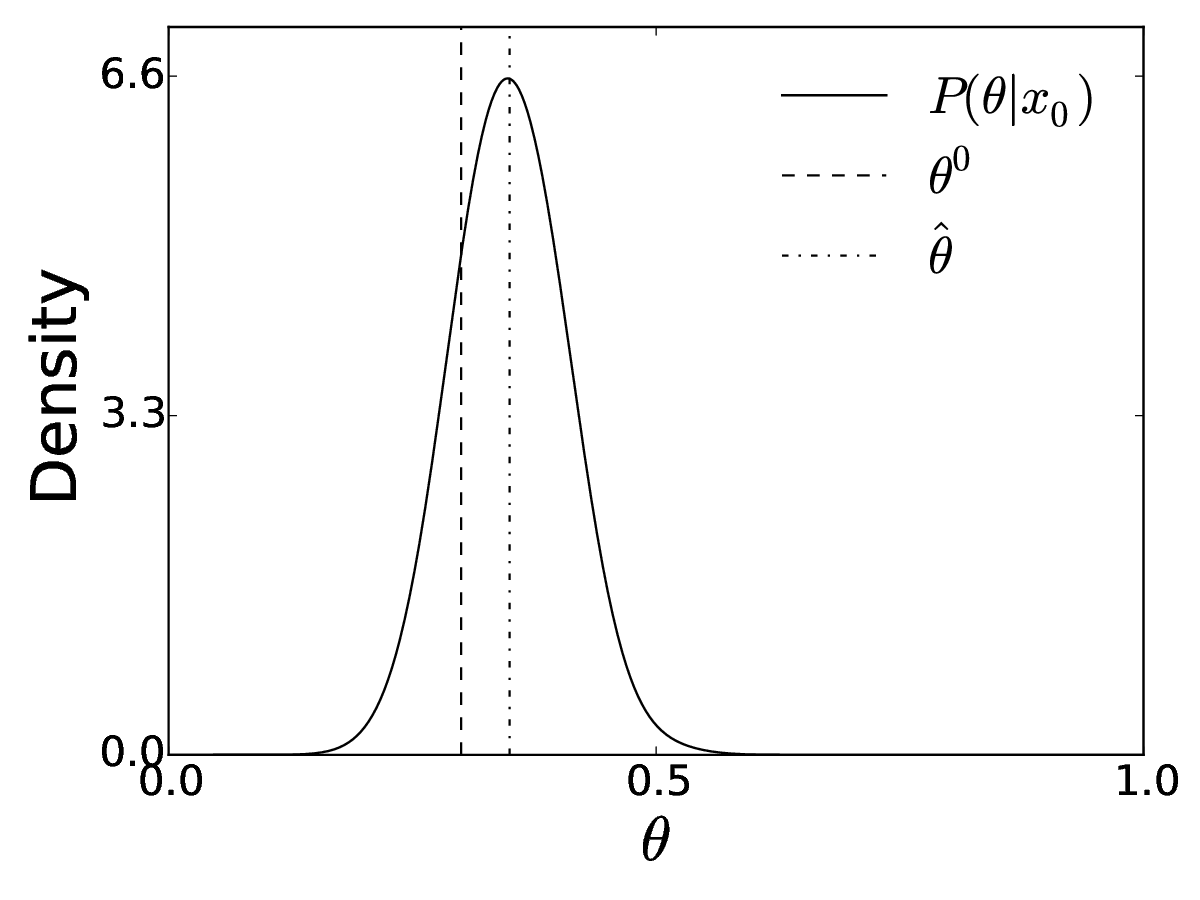}
    \label{fig:er_SC_theta_posterior}
    \adjustbox{valign=T}{\subfigure{e.}}
    \adjincludegraphics[valign=T,scale=0.2]{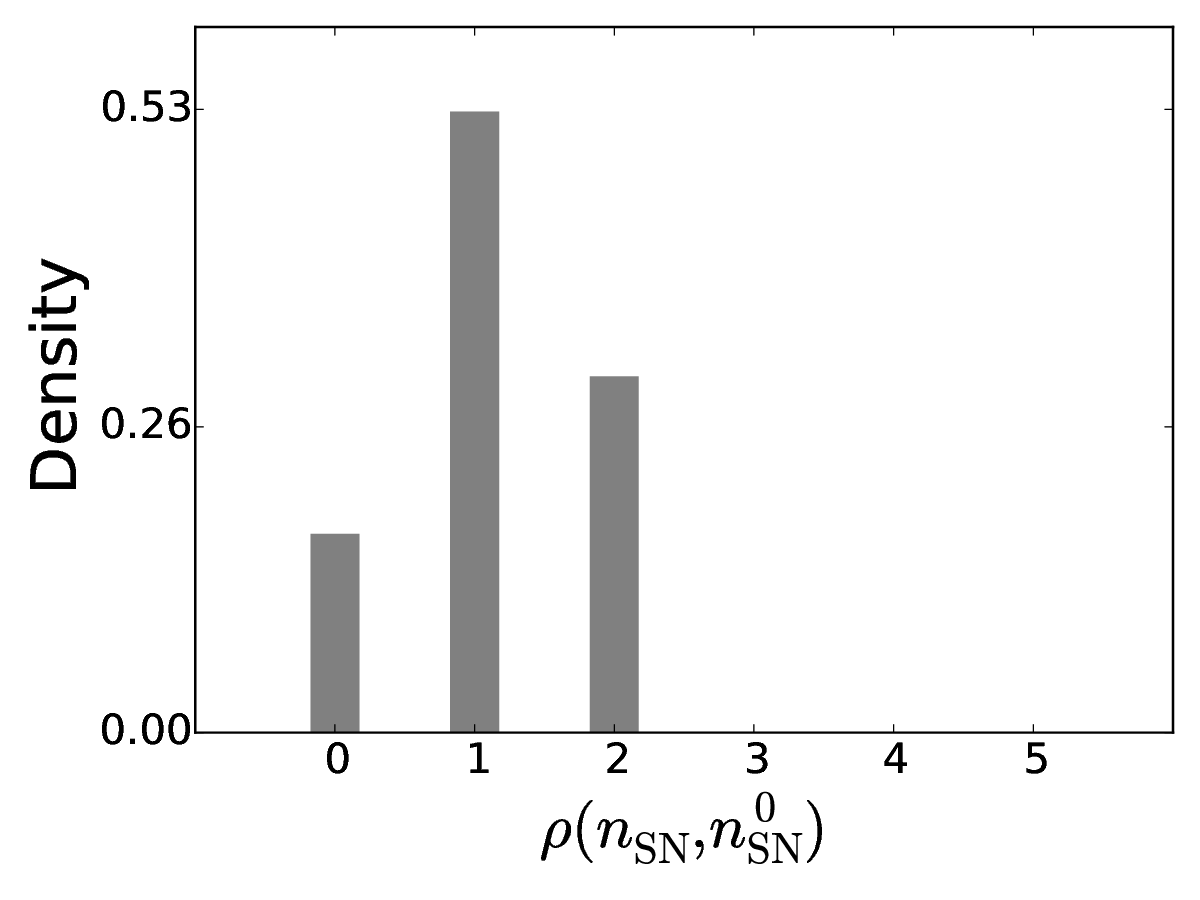}
    \label{fig:er_SC_seednode_posterior}
    \adjustbox{valign=T}{\subfigure{f.}}
    \adjincludegraphics[valign=T,scale=0.2]{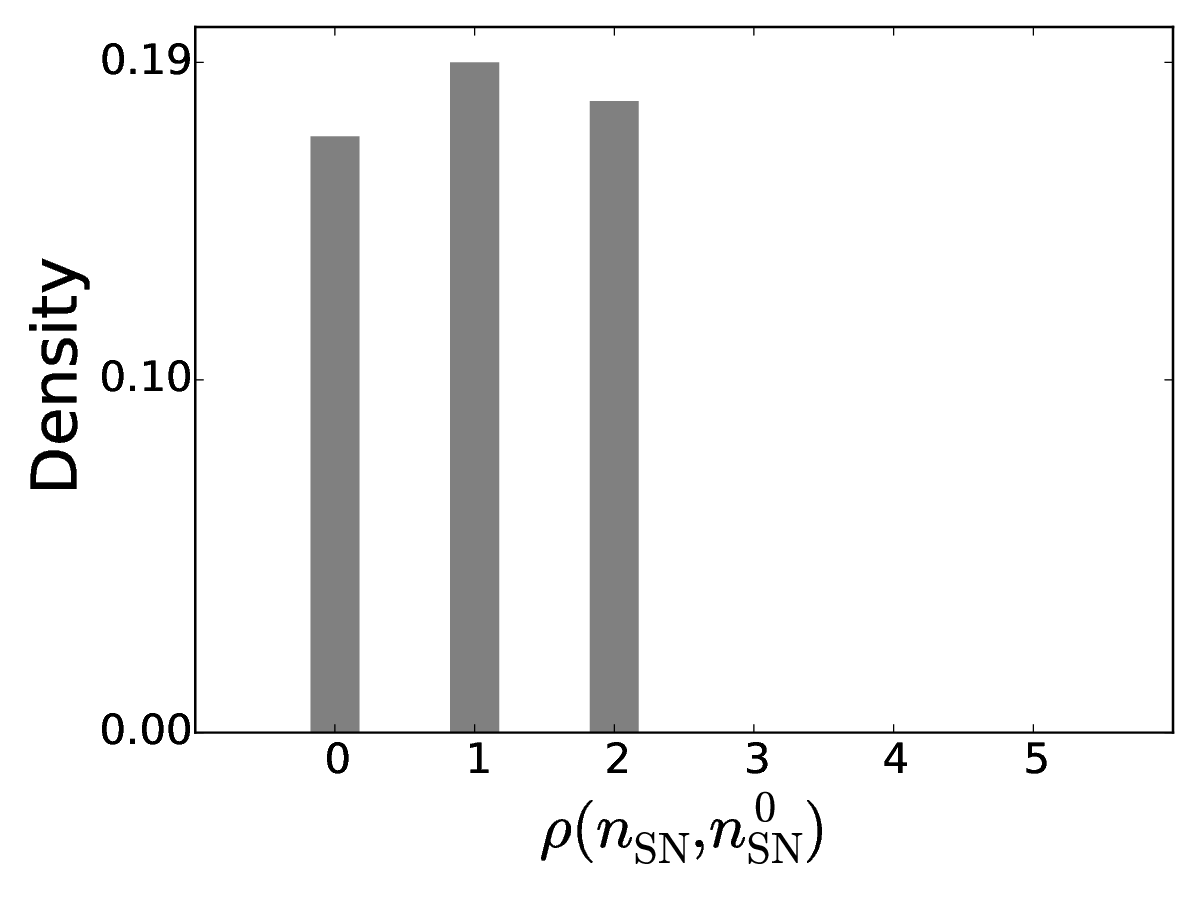}
    \label{fig:er_SC_seednode_posterior_max}
    
    \adjustbox{valign=T}{\subfigure{g.}}
    \adjincludegraphics[valign=T,scale=0.2]{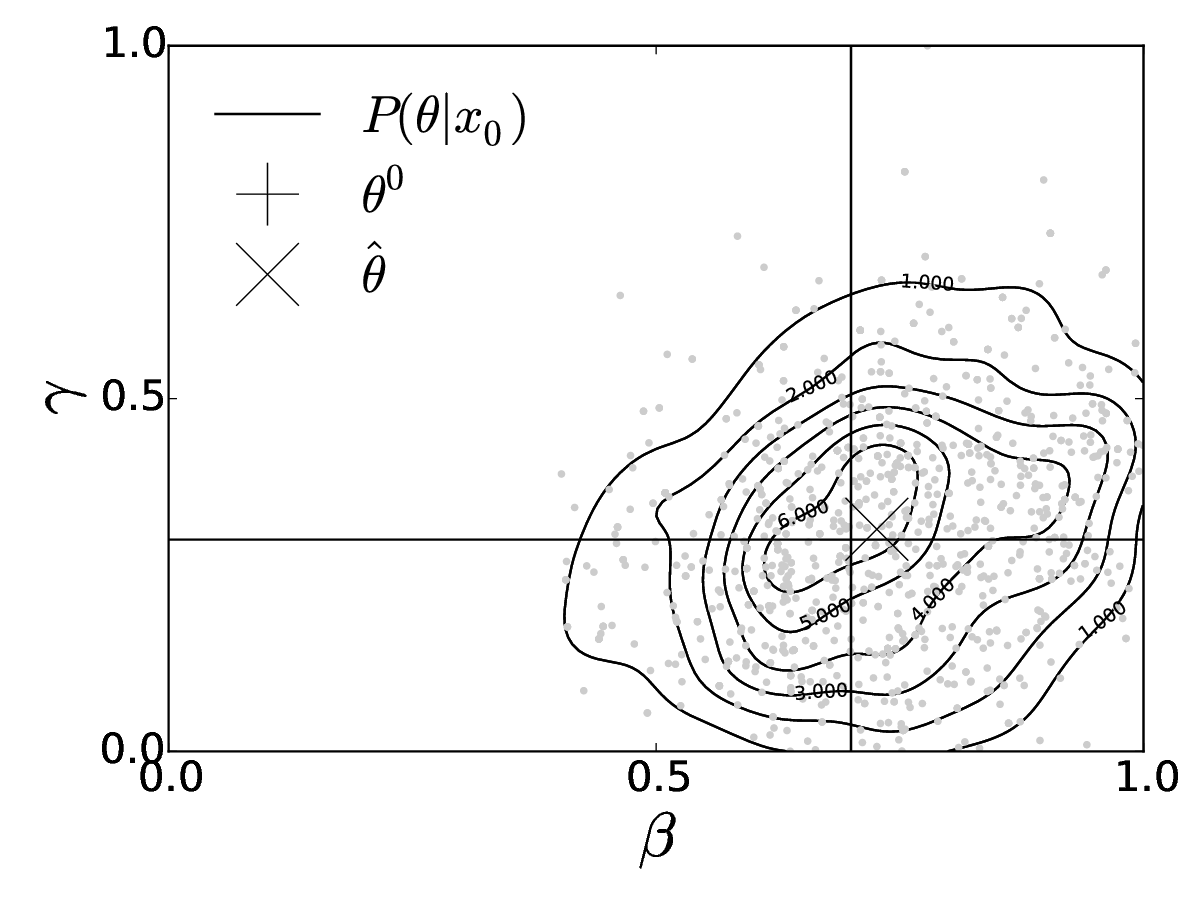}
    \label{fig:ba_SC_theta_posterior}
    \adjustbox{valign=T}{\subfigure{h.}}
    \adjincludegraphics[valign=T,scale=0.2]{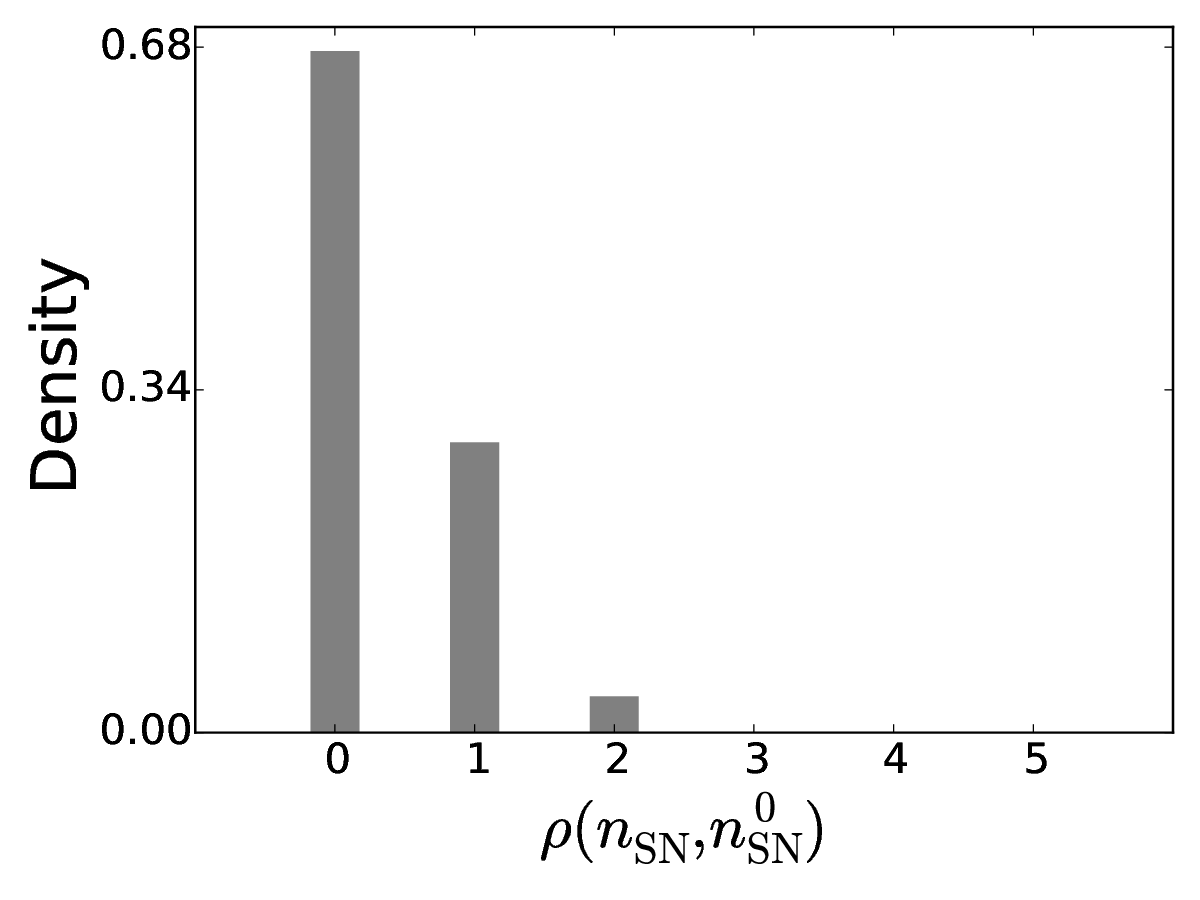}    
    \label{fig:ba_CC_beta_gamma_posterior}    
    \adjustbox{valign=T}{\subfigure{i.}}
    \adjincludegraphics[valign=T,scale=0.2]{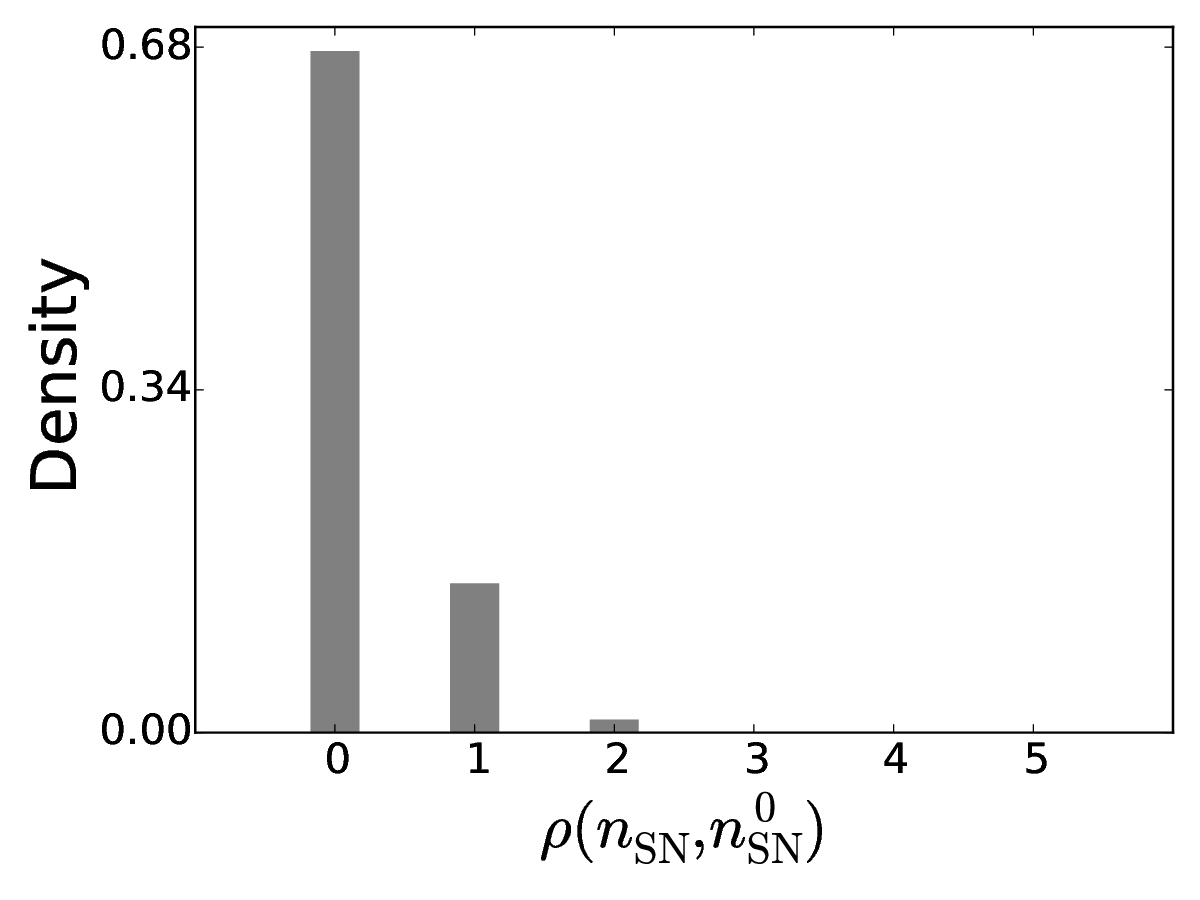}    
    \label{fig:ba_CC_beta_gamma_posterior_max}    
        
    \adjustbox{valign=T}{\subfigure{j.}}
    \adjincludegraphics[valign=T,scale=0.2]{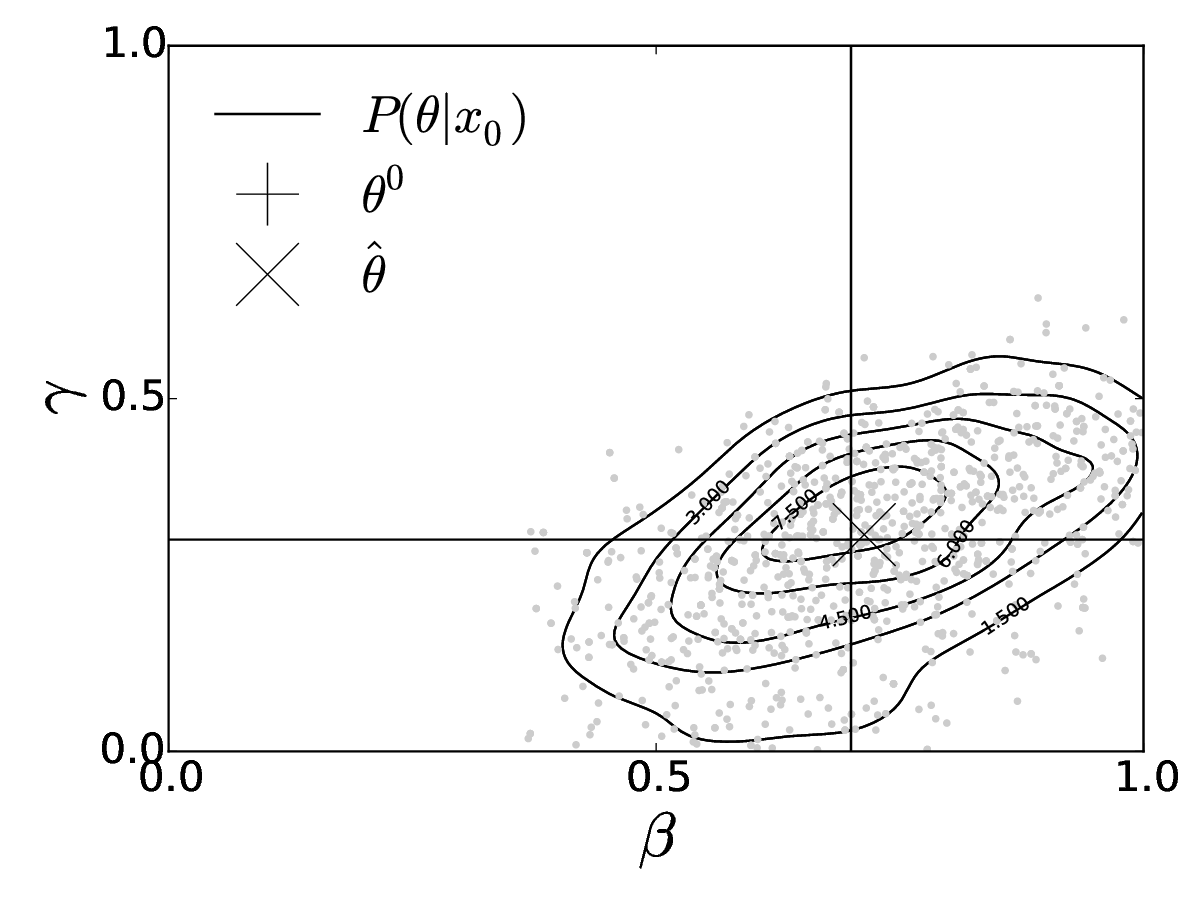}
    \label{fig:ba_CC_seednode_posterior}
    \adjustbox{valign=T}{\subfigure{k.}}
    \adjincludegraphics[valign=T,scale=0.2]{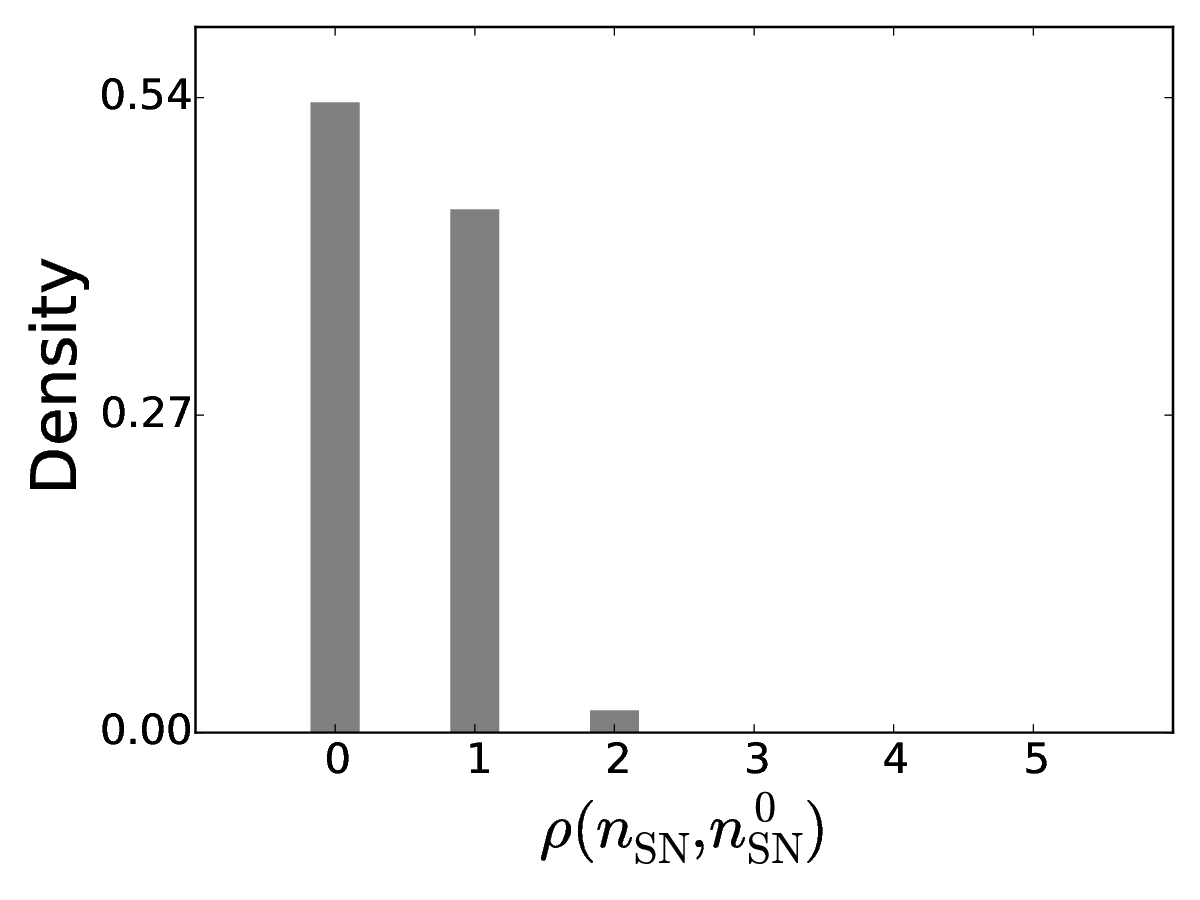}
    \label{fig:er_CC_beta_gamma_posterior}
    \adjustbox{valign=T}{\subfigure{l.}}
    \adjincludegraphics[valign=T,scale=0.2]{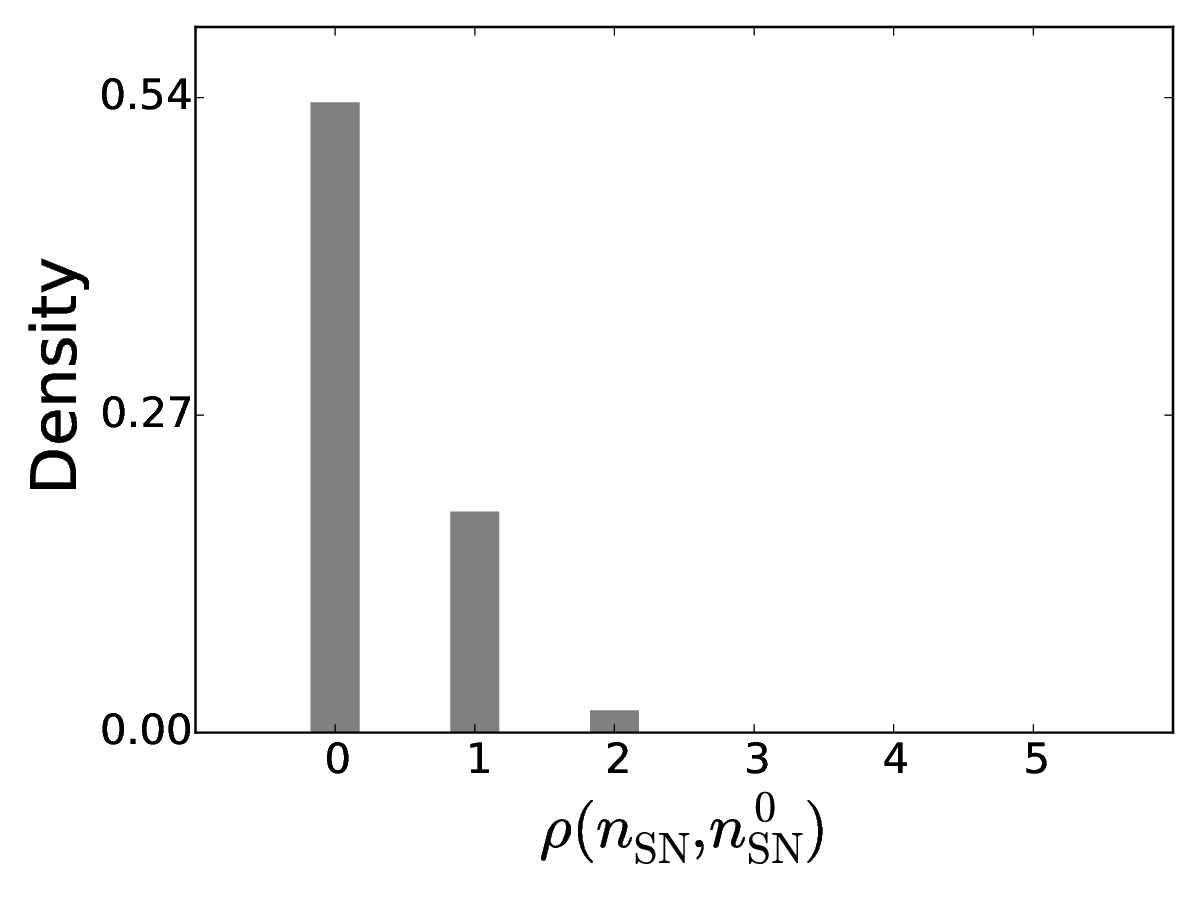}
    \label{fig:er_CC_beta_gamma_posterior}
\caption{\textbf{Inference of simple and complex contagion models.} Density of the inferred marginal posterior distribution of parameters of the spreading process and $\seednode$, given $\dataObs$, for epidemics on \ba (BA) and \er (ER) networks. The leftmost panels (\textbf{a,d,g,j}) describe properties of the spreading process whereas the middle and rightmost panels (\textbf{b,e,h,k,c,f,i,l}) deal with seed-node identification. For the simple contagion, we show the marginal posterior distribution of $\ratediffusion$ in panels 
\textbf{a} and \textbf{d}, where the true values are $\ratediffusion^0 = 0.3$. The corresponding plots for complex contagion are shown in panels \textbf{g} and \textbf{j}, the distribution of $\parameter$, where the true values are $\parameter^0 \equiv (\rateexposure^0, \threshold^0) = (0.7, 0.3)$.
Panels \textbf{b,e} and \textbf{c,f} display, respectively, the marginal posterior distribution and maximum posterior distribution on a single node at different distances from the true seed-node $\seednode^0$ for simple contagion model, while the corresponding plots for complex contagion process are in panels \textbf{h,k} and \textbf{i,l}. The density for $\ratediffusion$ and $\parameter$ were computed using Gaussian kernel density estimation on the samples from the posterior distribution \cite{Scott_2015}. 
Panels (\textbf{a,b,c,g,h,i}) refer to BA; (\textbf{d,e,f,j,k,l}) refer to ER. The diameters of the simulated BA and ER networks are, respectively, 4 and 6.}
  \label{fig:posterior_joint}
  \end{figure}

1000 IID samples were drawn from the posterior distribution of $\parametertheo$, $p(\parametertheo|\dataObs)$, using SABC inference scheme. The inferred marginal posterior distribution of $\parameter$ and $\seednode$ on the BA and ER networks are shown in Figure~\ref{fig:posterior_joint} for the simple and complex contagion processes. We recover the true values of the spreading parameter $\parameter^0$ (dashed vertical line), which has a high posterior probability (left panels). The posterior distribution of the seed-node (right panels) is also concentrated in the network neighborhood of $\seednode^0$.    

\subsection*{Bayes estimates}
\label{sec:extimates}
As the contagion models are stochastic, $\dataObs$ generated from contagion model with the same values of $\parametertheo^0$ will also be a random quantity. Hence $\estparametertheo = (\estparameter, \estseednode)$, being a function of the observed dataset $\dataObs$, is also random. To quantify the randomness of the Bayes estimator, we generated 100 observed datasets as before, using both simple and complex contagion, on the BA and ER networks for the same parameter values. We infer the posterior distribution and the Bayes estimate using Equation~\ref{eq:Bayes_estimate} for each dataset.

\begin{figure}[htbp]
    \centering
    \mbox{}
    \adjustbox{valign=T}{\subfigure{a.}}
    \adjincludegraphics[valign=T,scale=0.3]{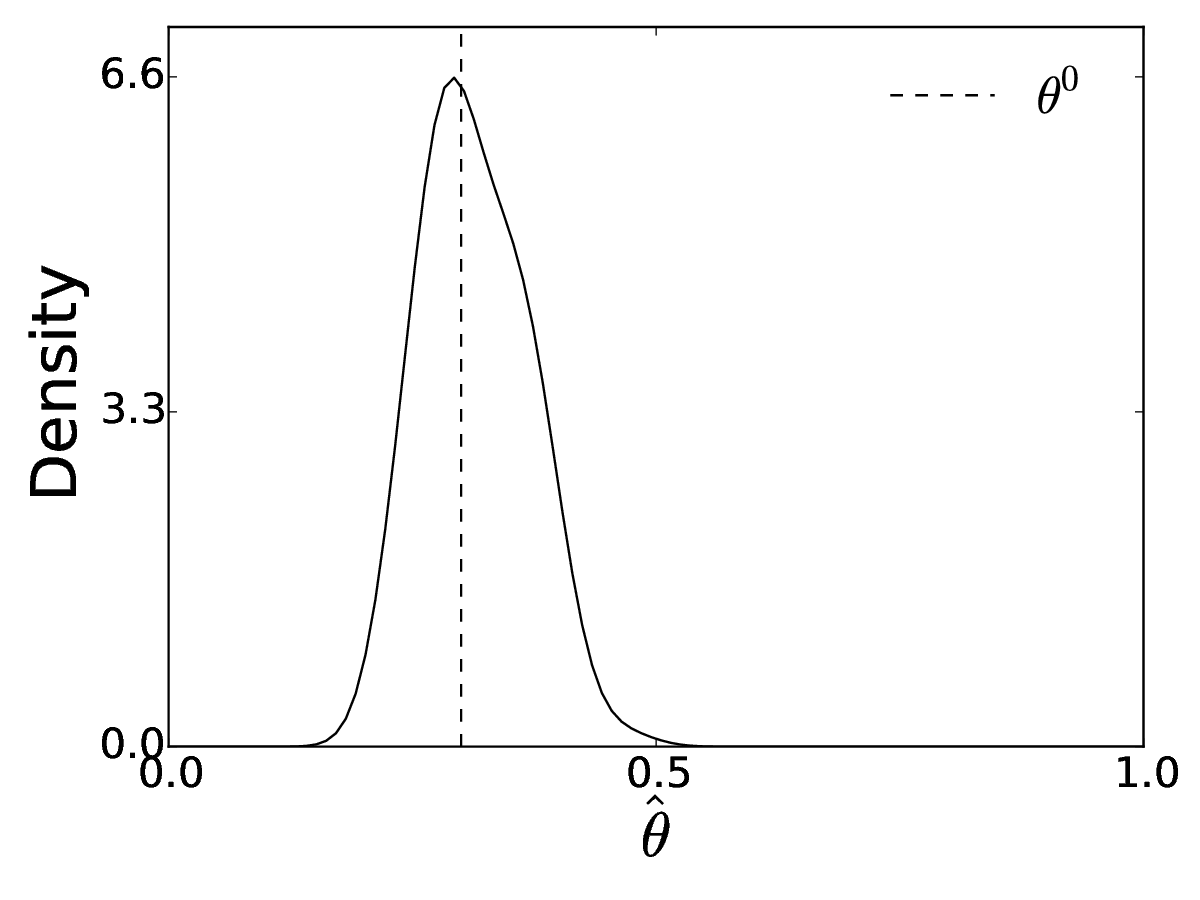}
    \label{fig:ba_SC_theta_est}
    \adjustbox{valign=T}{\subfigure{b.}}
    \adjincludegraphics[valign=T,scale=0.3]{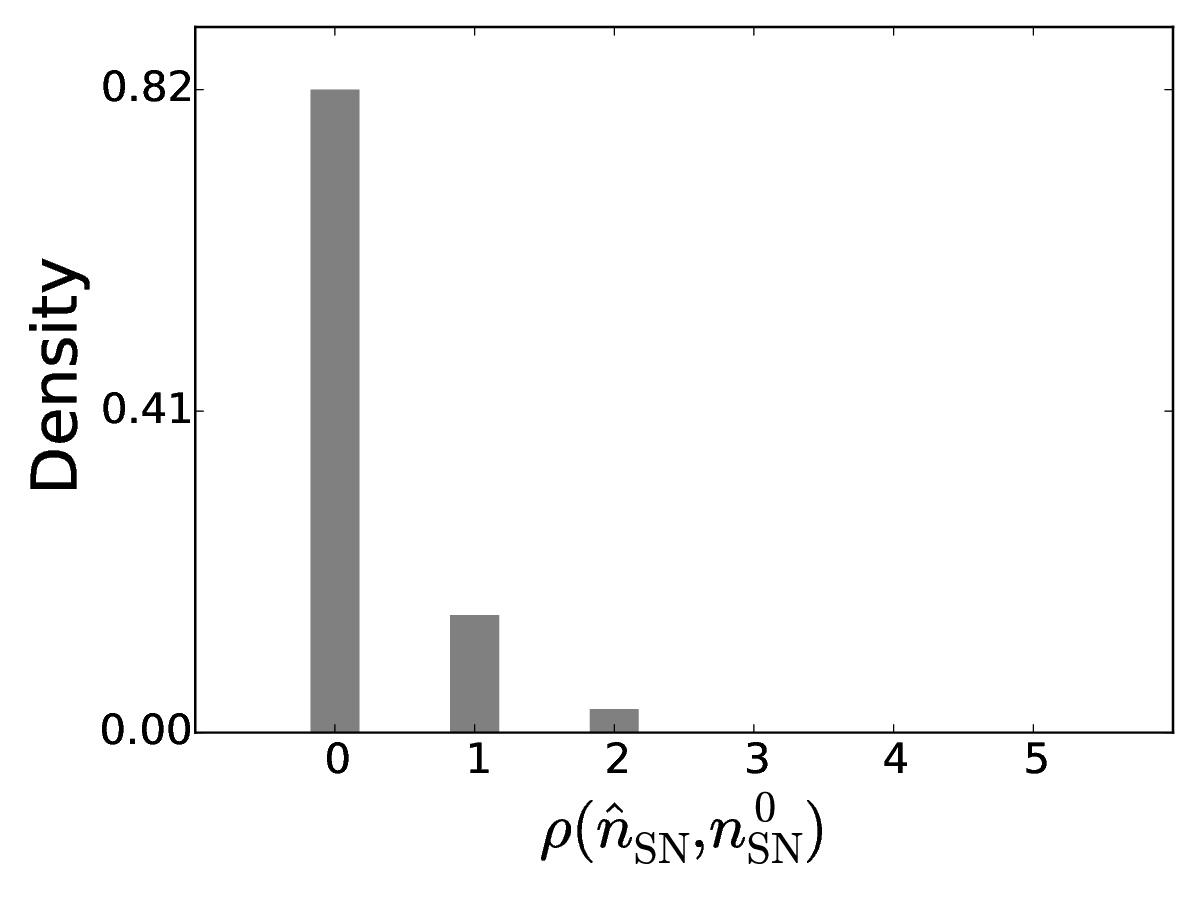}
    \label{fig:ba_SC_seednode_est}
    
    \adjustbox{valign=T}{\subfigure{c.}}
    \adjincludegraphics[valign=T,scale=0.3]{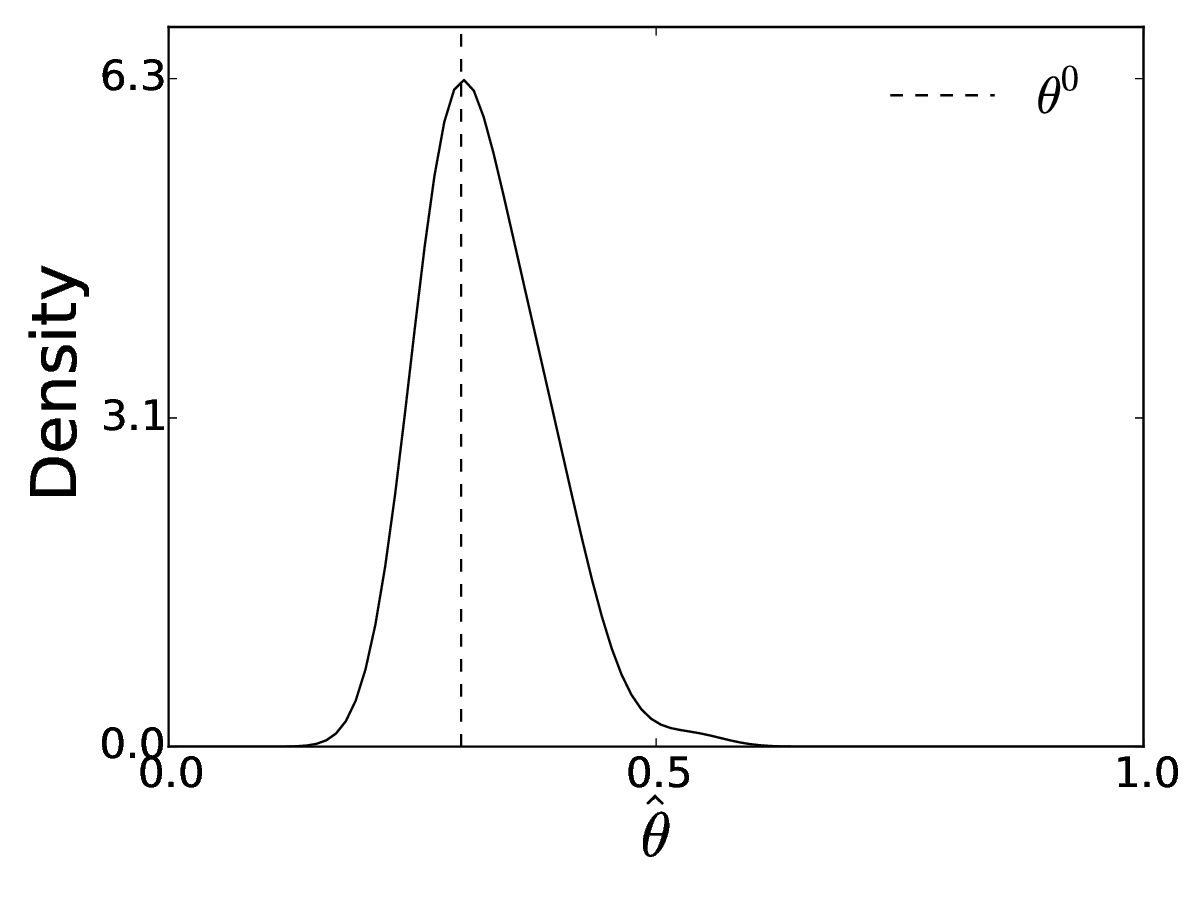}
    \label{fig:er_SC_theta_est}
    \adjustbox{valign=T}{\subfigure{d.}}
    \adjincludegraphics[valign=T,scale=0.3]{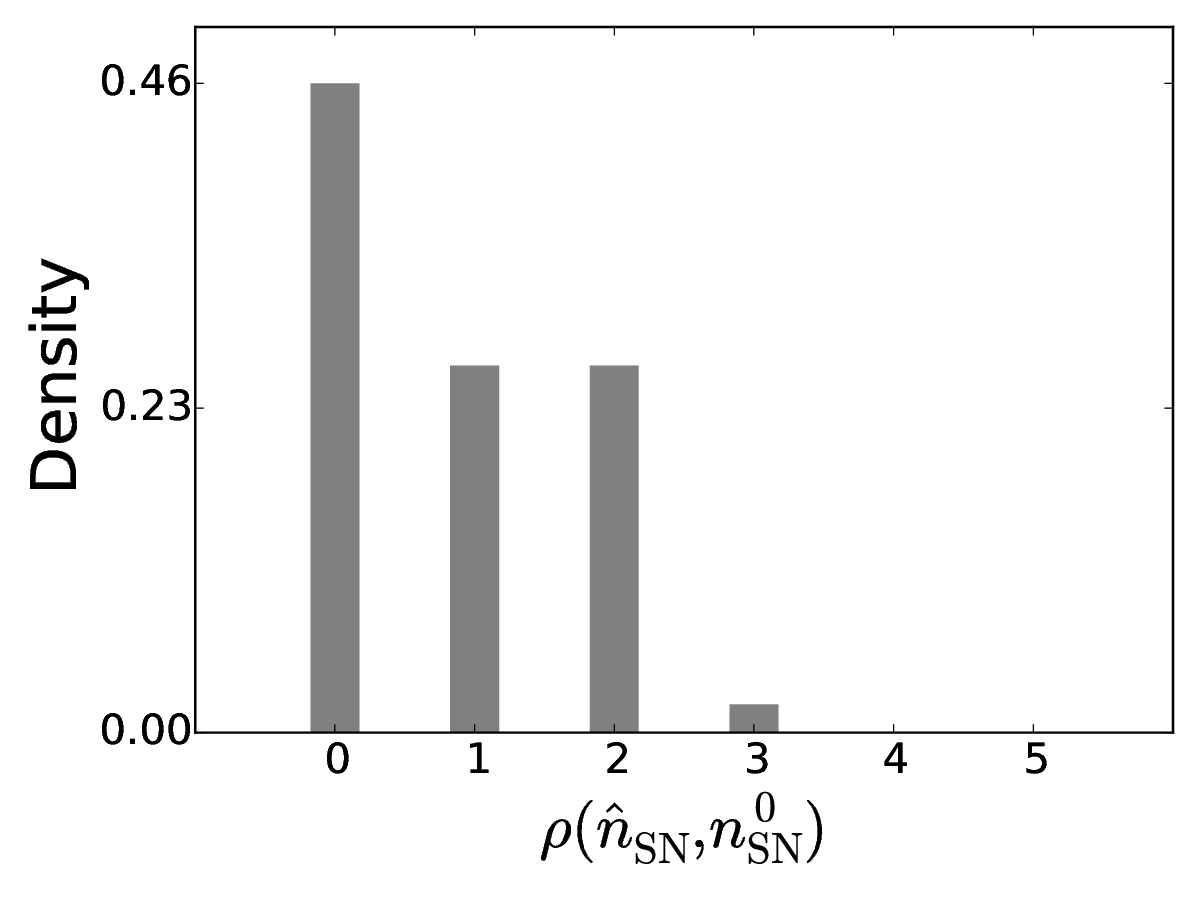}
    \label{fig:er_SC_seednode_est}
    
        \adjustbox{valign=T}{\subfigure{e.}}
    \adjincludegraphics[valign=T,scale=0.3]{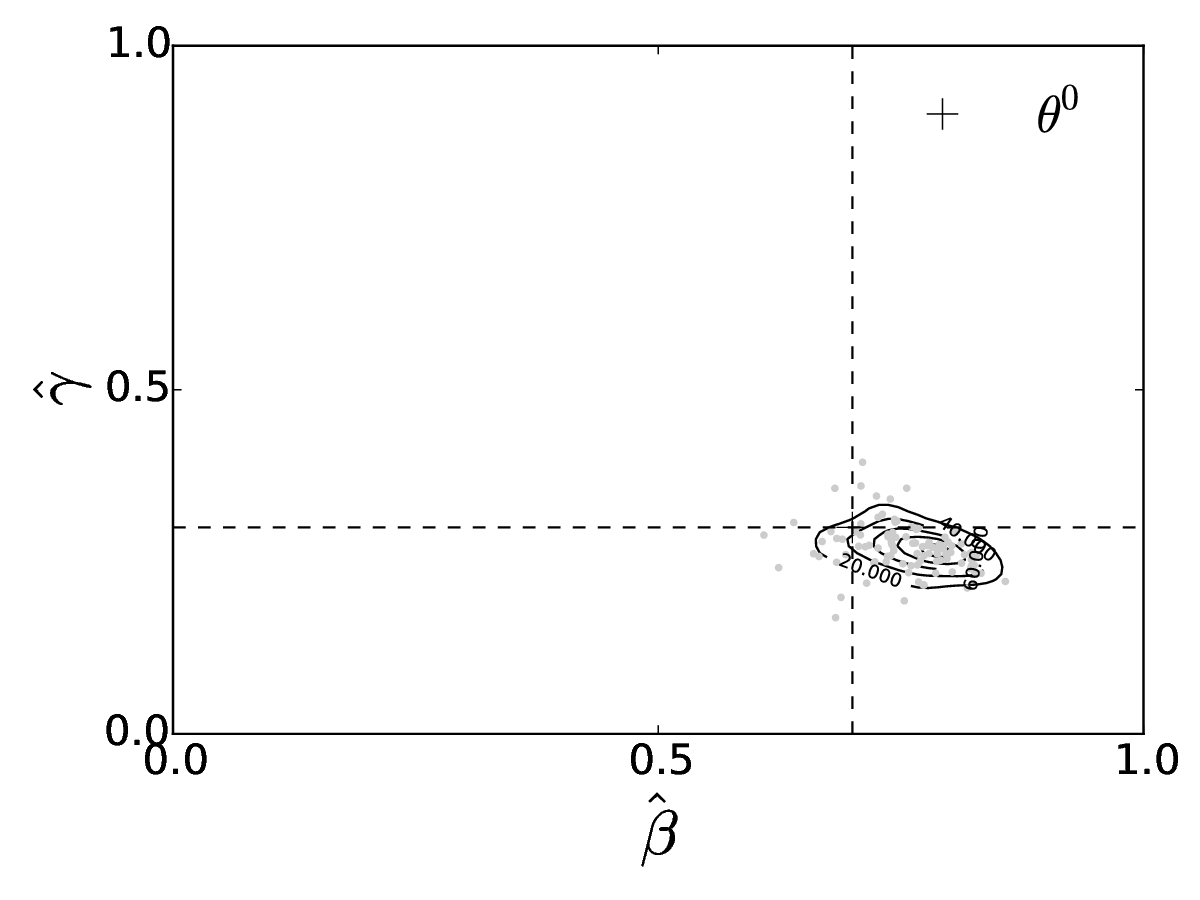}
    \label{fig:ba_CC_beta_gamma_est}
    \adjustbox{valign=T}{\subfigure{f.}}
    \adjincludegraphics[valign=T,scale=0.3]{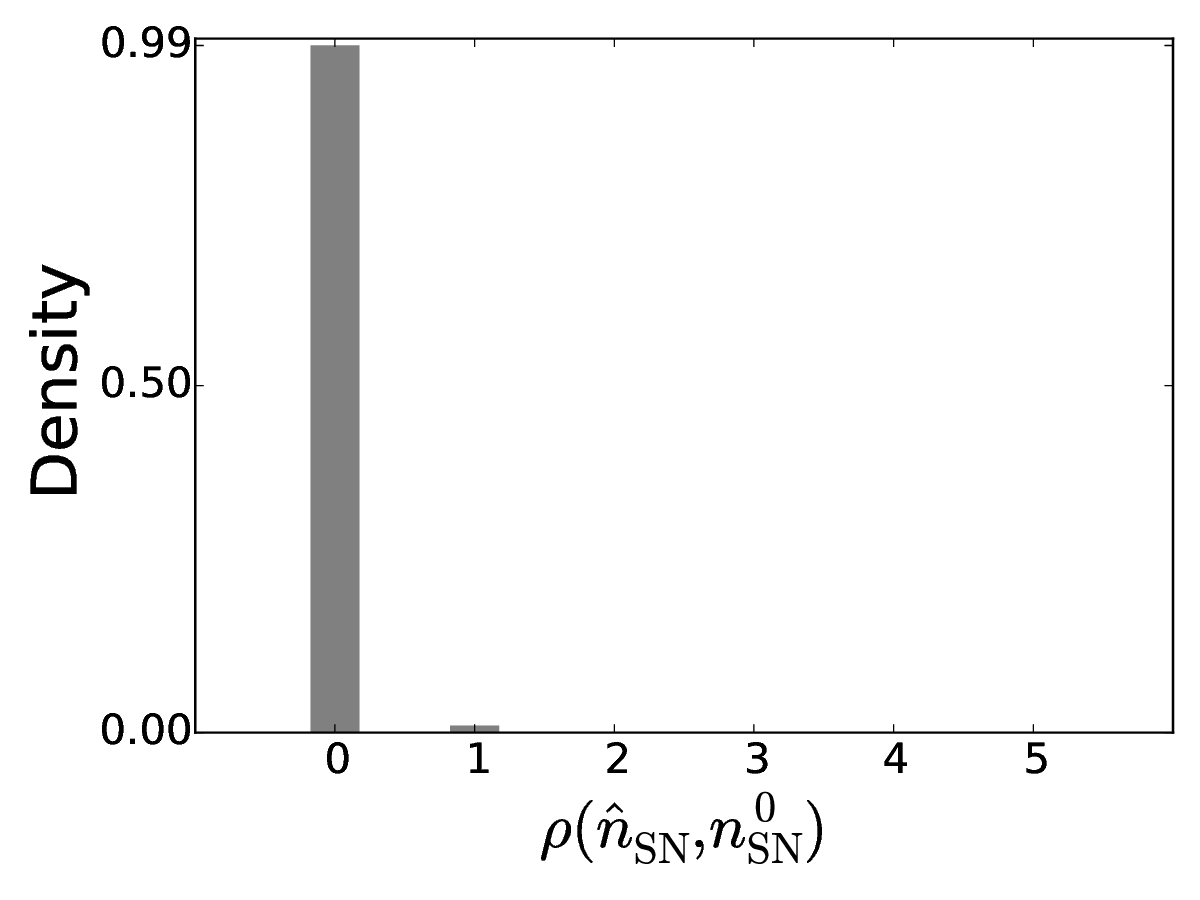}
    \label{fig:ba_CC_seednode_bias}
    
    \adjustbox{valign=T}{\subfigure{g.}}
    \adjincludegraphics[valign=T,scale=0.3]{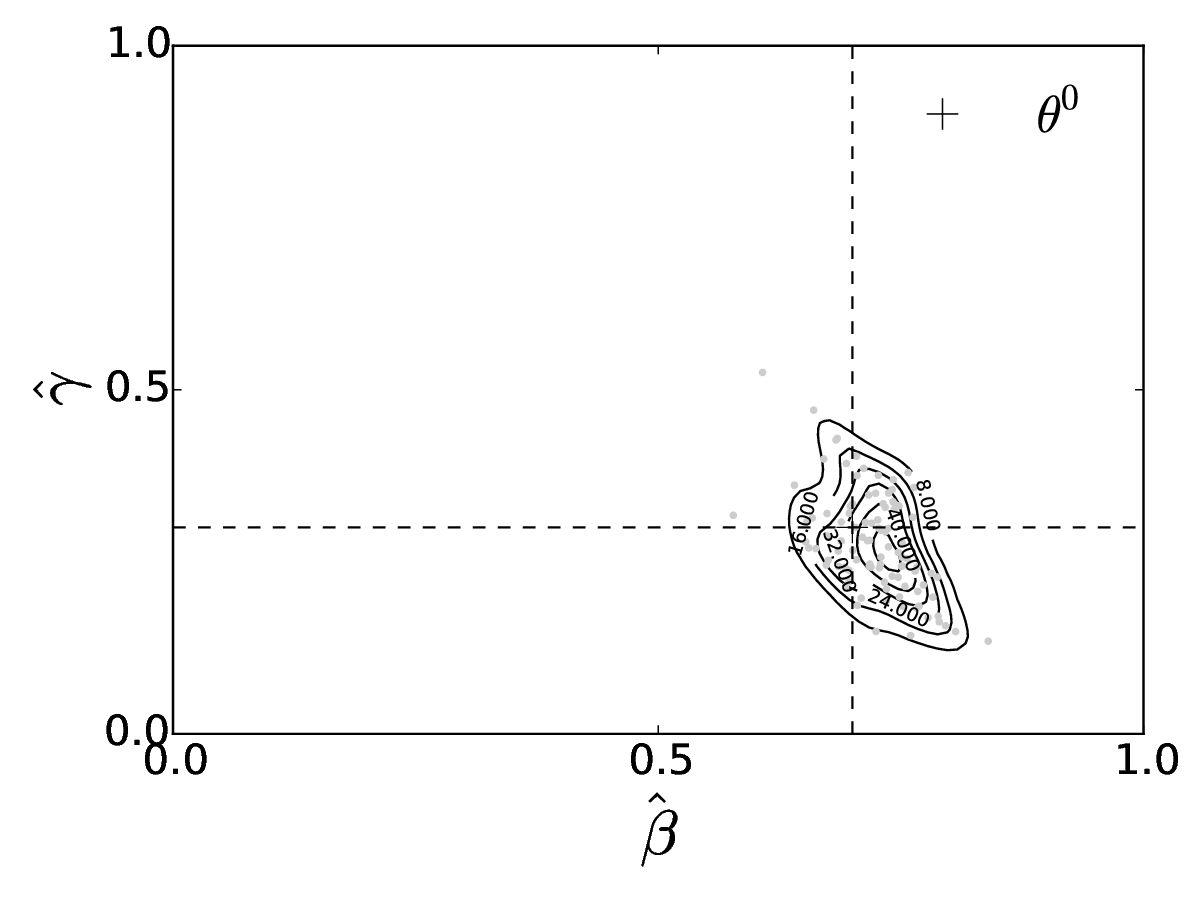}
    \label{fig:er_CC_beta_gamma_est}
    \adjustbox{valign=T}{\subfigure{h.}}
    \adjincludegraphics[valign=T,scale=0.3]{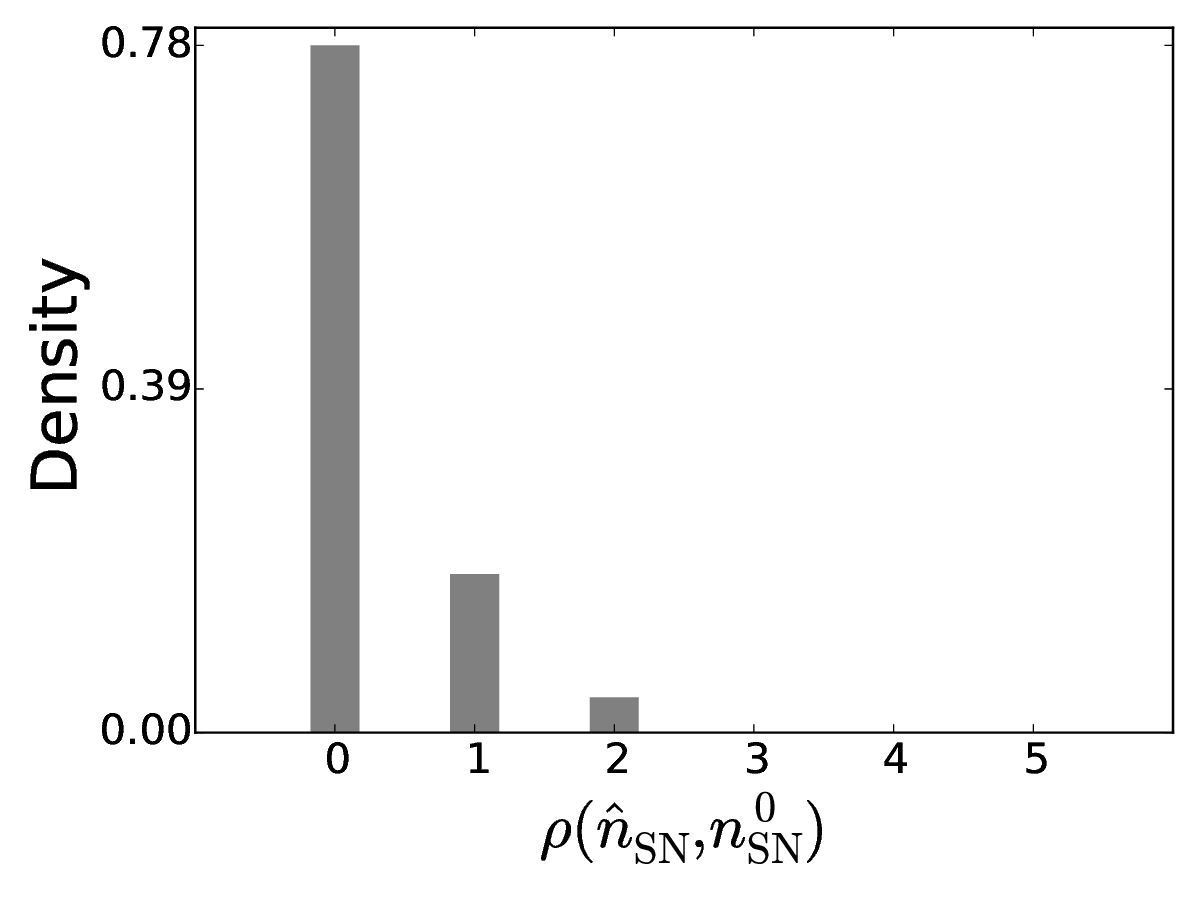}
    \label{fig:er_CC_seednode_est}
\caption{\textbf{Parameter estimates of simple and complex contagion models.} For the simple contagion model, we display the marginal densities of $\estratediffusion$ and marginal density of $\estseednode$ at different distances from the true seed-node $\seednode^0$, estimated from 100 observed datasets $\dataObs$ on the BA (panels \textbf{a,b}) and ER networks (panels \textbf{c,d}). For the complex contagion model, the corresponding results are shown for the BA model (panels \textbf{e,f}) and the ER model (panels \textbf{g,h}). To estimate the densities of $\estratediffusion$ and $\estparameter$, we used Gaussian kernel density estimation with band-width 0.6. The diameters of the simulated BA and ER networks are, respectively, 4 and 6.
}
\label{fig:Bayes_est}
  \end{figure}

The marginal density of the estimated parameters from 100 different realizations of $\dataObs$ of simple and complex contagion model on BA and ER networks are shown in Figure~\ref{fig:Bayes_est}. We note that the density of $\estseednode$ peaks at $\seednode^0$, the true source node, with a significant concentration on the nodes that are at path lengths $\leq 2$ from $\seednode^0$. Similarly, the estimates of the parameters of the spreading processes are concentrated in a neighborhood of the true value, showing the robustness of the parameter estimation scheme for different epidemic models and different synthetic network topologies.  

\subsection*{Sensitivity of inference to the number of observations}
\label{sec:sensitivity}
\begin{figure}[htbp]
    \centering
    \mbox{}%
    \adjustbox{valign=T}{\subfigure{a.}}
    \adjincludegraphics[valign=T,scale=0.39]{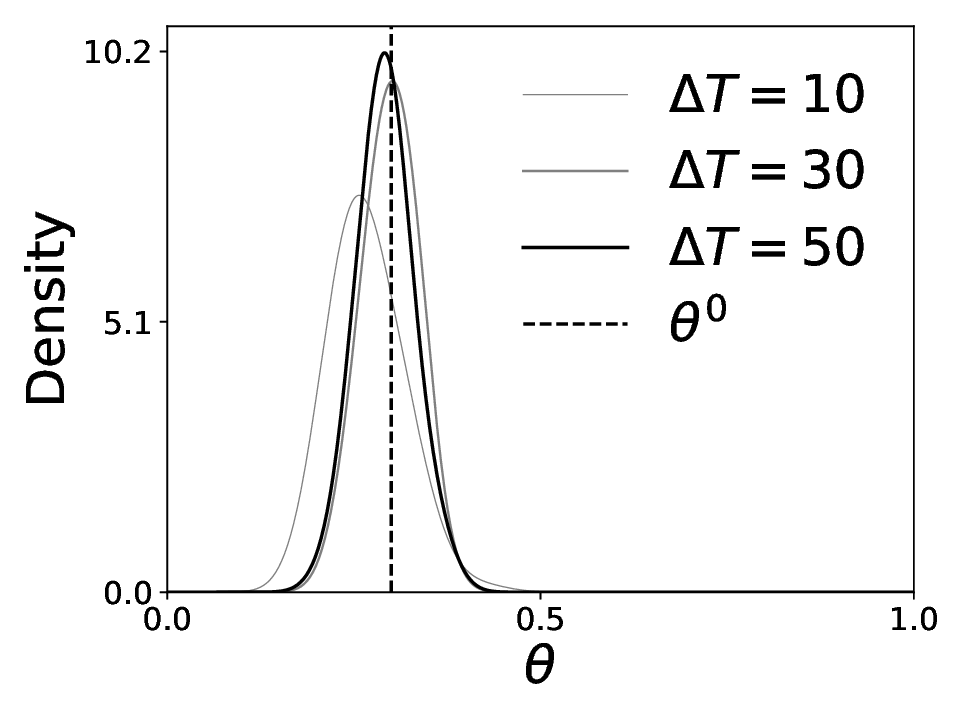}
    \label{fig:ba_time_dependence_theta}
    \adjustbox{valign=T}{\subfigure{b.}}
    \adjincludegraphics[valign=T,scale=0.39]{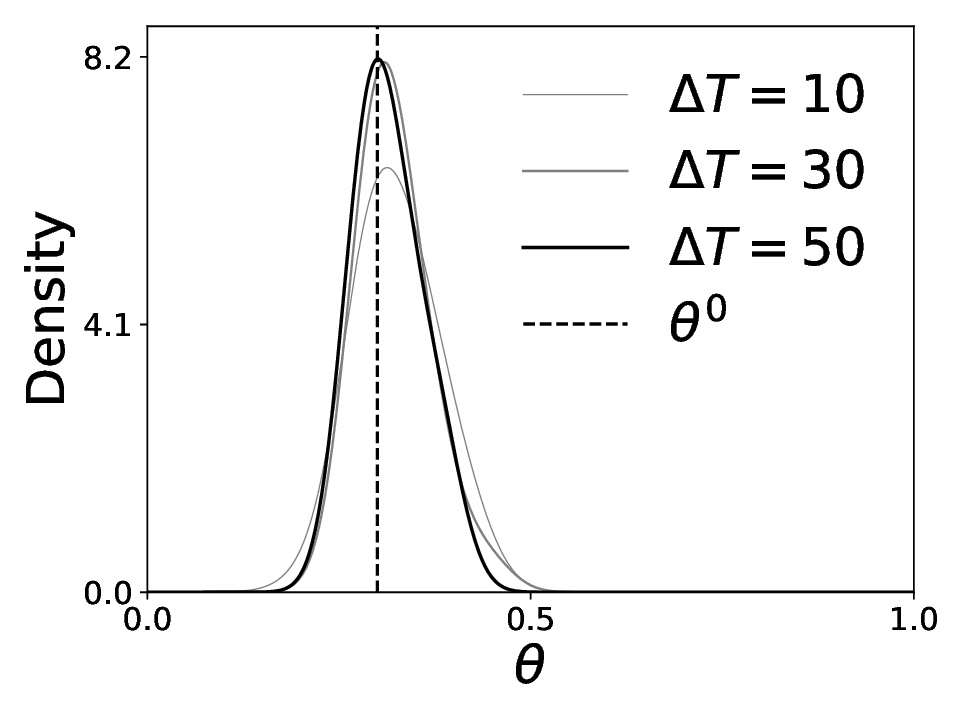}
    \label{fig:er_time_dependence_theta}
    
    \adjustbox{valign=T}{\subfigure{c.}}    
    \adjincludegraphics[valign=T,scale=0.39]{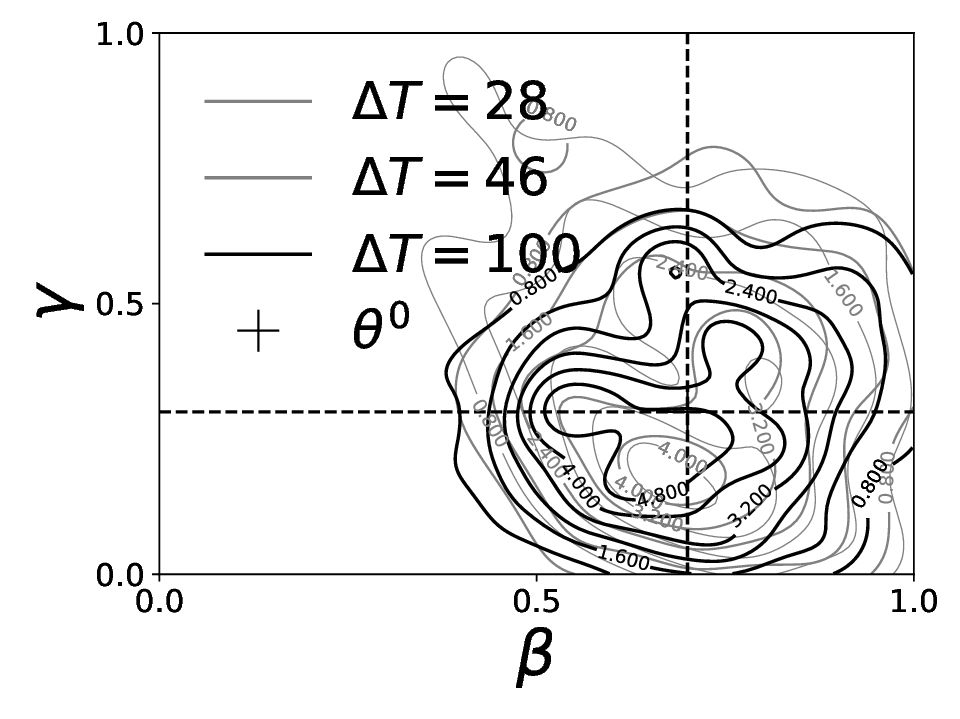}
    \label{fig:ba_time_dependence_beta_gamma}
    \adjustbox{valign=T}{\subfigure{d.}}
    \adjincludegraphics[valign=T,scale=0.39]{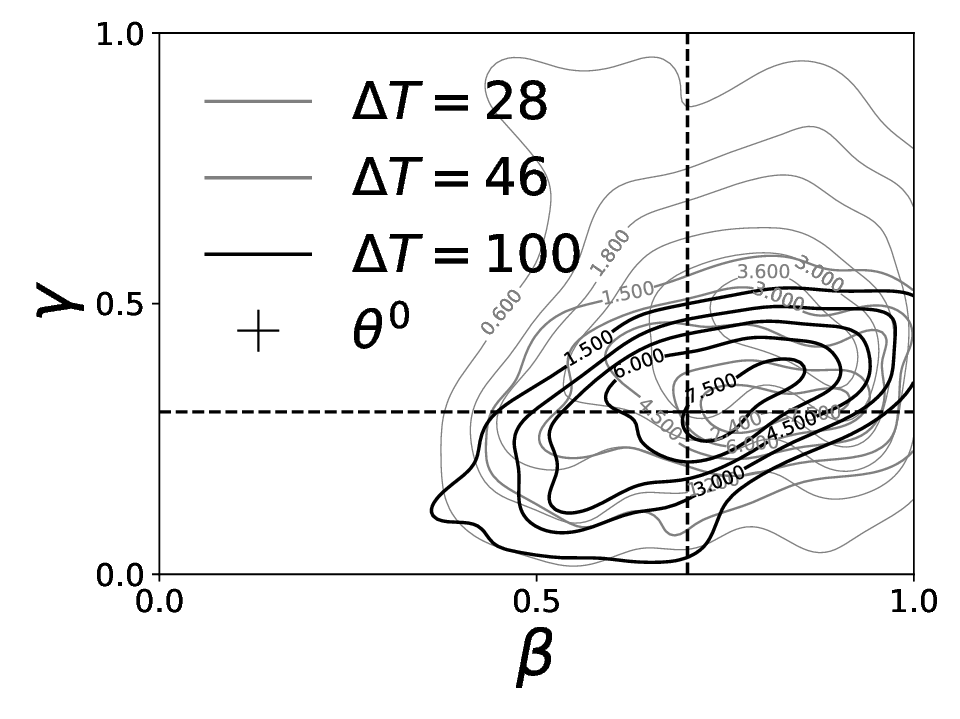}
    \label{fig:er_time_dependence_beta_gamma}
\caption{\textbf{Dependence of inference on $\Delta T = T-t_0$.} Density of the inferred marginal posterior distribution of $\ratediffusion$ and $\parameter$, given $\Delta T = T-t_0$ observed time steps in $\dataObs$ starting at $t_0=20$, the simple (panels \textbf{a, b}) and complex (panels \textbf{c, d}) contagion process simulated on the Barbasi-Albert (BA) and Erd\"os-R\`enyi (ER) networks correspondingly. (\textbf{a,c}): BA; (\textbf{b,d}):ER. The diameters of the simulated BA and ER networks are, respectively, 4 and 6.}
\label{fig:posterior_timedependence}
  \end{figure}

In the cases considered $\dataObs$ contained $\Delta T = 50$ and $\Delta T = 100$ observed time steps for simple and complex contagion models, respectively, where $\Delta T = T-t_0$. 
Next, we explored how different values of $\Delta T$ influence the accuracy of inference. Inference of the seed-node only depends on the states of the nodes of the network observed at $t = t_0$ and is unaffected by the observations at subsequent time steps. For the sensitivity study, we therefore only consider inference of the remaining model parameters. In Figure~\ref{fig:posterior_timedependence}, we show the inferred marginal posterior distribution of model parameters ($\ratediffusion$ and $\parameter$ respectively for simple and complex contagion) when considering only the first $\Delta T$ time steps of the observed dataset for different values of $\Delta T$. The posterior distribution is concentrated in a neighborhood of the true value for all values of $\Delta T$, though we notice, as expected, that the concentration of the posterior distribution around the true value increases with the observed number of time steps ($\Delta T$).
We conclude that our inference scheme can perform well also with a small number of observed time steps ($\Delta T$), though the performance improves with the increase of $\Delta T$, which is intuitive as each observed time step carries additional information on the unknown value of the model parameter that has generated the observed dataset $\dataObs$.

\subsection*{Comparison with the Netsleuth algorithm}

One of the outcomes of the suggested parameter estimation scheme is the estimation of a seed-node, which has recently received attention in modeling the spread of disease or information on social networks \cite{lappas2010finding, shah2011rumors, prakash2012spotting}. As the likelihood function for the source node initiating a general spreading process on a general network topology is intractable, existing research on source node detection assumes a specific contagion process and a fixed network topology, typically a tree. Under these assumptions, one estimates the source node by maximizing an approximate likelihood function and then proposes a heuristic algorithm for doing inference on a general network topology by assuming the approximation to hold more generally. In the spirit of indirect inference \cite{drovandi2011approximate} we could use the estimates of the seed-node and contagion probability from these simple models available in the literature as summary statistics in our ABC framework. Note that, crucially, to this aim there is no need for a closed form expression of the likelihood of these simple models unlike what is typically advocated for indirect inference, see for example in \cite{drovandi2015bayesian}. The different contributions in this literature vary depending on what type of spreading  process is assumed to operate on the network. Processes that have been studied up to date include the independent cascade model \cite{lappas2010finding}, the susceptible-infected (SI) process \cite{shah2011rumors, luo2012identifying, prakash2012spotting, dong2013rooting, prakash2014efficiently, sundareisan2015hidden} which coincides with our simple contagion process, the susceptible-infected-recovered (SIR) \cite{zhu2016information}, the heterogeneous SIR \cite{zhu2014robust} where the approximate likelihood has been demonstrated to be asymptotically valid in geometric trees \cite{shah2011rumors}, and spreading processes on infinite trees \cite{zhu2016information}. All approaches we are aware of \emph{use spreading process specific approximations of the likelihood function and apply them to specific network topologies}. In contrast, in our approach the implicit approximation of the likelihood function does not depend on the specific spreading process or the specific network topology under consideration, but it does however depend on the choice of the ABC discrepancy measure between the observed and simulated data. We have given an intuitive justification behind the choice of the discrepancy measure in Section~\ref{sec:if}.

  \begin{figure}[htbp]
    \centering
    \mbox{}
    \adjustbox{valign=T}{\subfigure{a.}}
    \adjincludegraphics[valign=T,scale=0.3]{ba_SC_network_bias_sn}
    \label{fig:ba_SC_theta_est}
    \adjustbox{valign=T}{\subfigure{b.}}
    \adjincludegraphics[valign=T,scale=0.3]{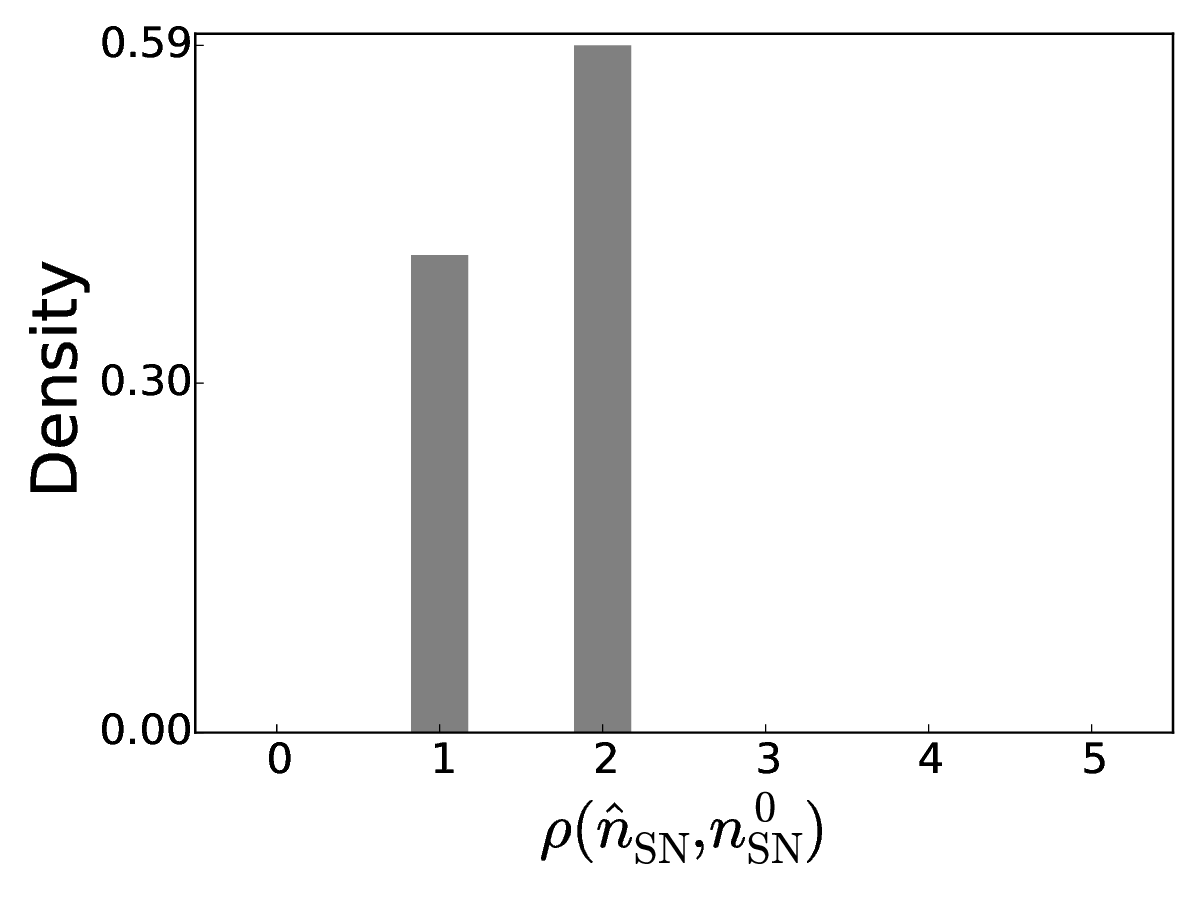}
    \label{fig:ba_SC_seednode_est}
    
    \adjustbox{valign=T}{\subfigure{c.}}
    \adjincludegraphics[valign=T,scale=0.3]{er_SC_network_bias_sn}
    \label{fig:er_SC_theta_est}
    \adjustbox{valign=T}{\subfigure{d.}}
    \adjincludegraphics[valign=T,scale=0.3]{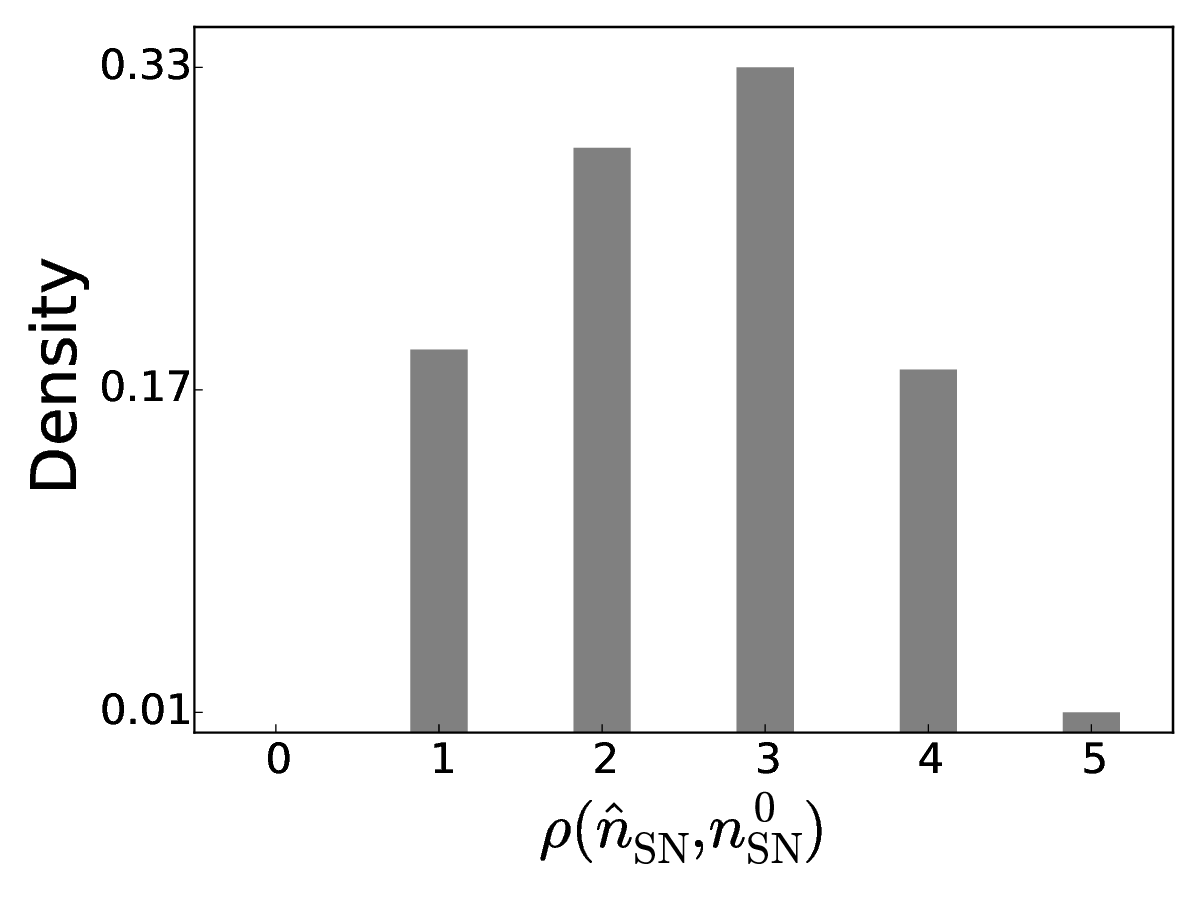}
    \label{fig:er_SC_seednode_est}
\caption{\textbf{Bayes and Netsleuth estimates of seed-node for simple contagion model.} For the simple contagion model, we display the marginal density of $\estseednode$ at different distances from the true seed-node $\seednode^0$, estimated from 100 observed datasets $\dataObs$ using proposed Bayesian scheme (panels \textbf{a,c}) and Netsleuth (panels \textbf{b,d}). (\textbf{a,b}): BA; (\textbf{c,d}):ER. The diameters of the simulated BA and ER networks are, respectively, 4 and 6.}
\label{fig:Comparison_est}
  \end{figure}
  
In Figure~\ref{fig:Comparison_est}, we compare our Bayes estimate of seed-node with estimates computed by the Netsleuth algorithm of  \cite{prakash2012spotting, prakash2014efficiently} for 100 different simple contagion process datasets. To find the estimate of the seed-node, the Netsleuth algorithm maximizes an approximate likelihood function under the assumption of a regular tree network topology for SI process. As the inference of the seed-node only depends on the observations at $t=\obsstarttime$, to infer the seed-node using the Netsleuth algorithm, we use the state of the nodes at $\obsstarttime$ with the true value of $\ratediffusion$ assumed to be known. Our simulation shows that similar to our methodology, the Netsleuth algorithm also performs better on more heterogeneous network topologies. However, the Bayes estimate is superior in detecting the true seed-node. We also note that the proposed Bayesian scheme is able to quantify uncertainty of the seed-node, while the Netsleuth algorithm only provides a point estimate.
 
The proposed method is substantially more expensive than the Netsleuth algorithm. 
NetSleuth algorithm and the method we propose, take respectively 0.002 (on a single core) and 391 seconds (using 30 nodes with 36 cores of Piz Daint Cray
architecture (Intel Broadwell + NVidia TESLA P100)), to infer the seed-node of an epidemic on simulated BA network with 100 nodes.
The most expensive step of Netsleuth algorithm is an eigenvalue decomposition, whereas SABC inference scheme depends on repeated simulations of datasets for different parameter values, making it computationally expensive. 
The reported inference results have been achieved by using the efficient HPC architecture developed in \cite{Dutta_2017_PASC}. 
 
\subsection{Inference on empirical networks}
\label{sec:emp_network}
\begin{figure}[htbp]
    \centering
    \mbox{}%
    \adjustbox{valign=T}{\subfigure{a.}}
    \adjincludegraphics[valign=T,scale=0.2]{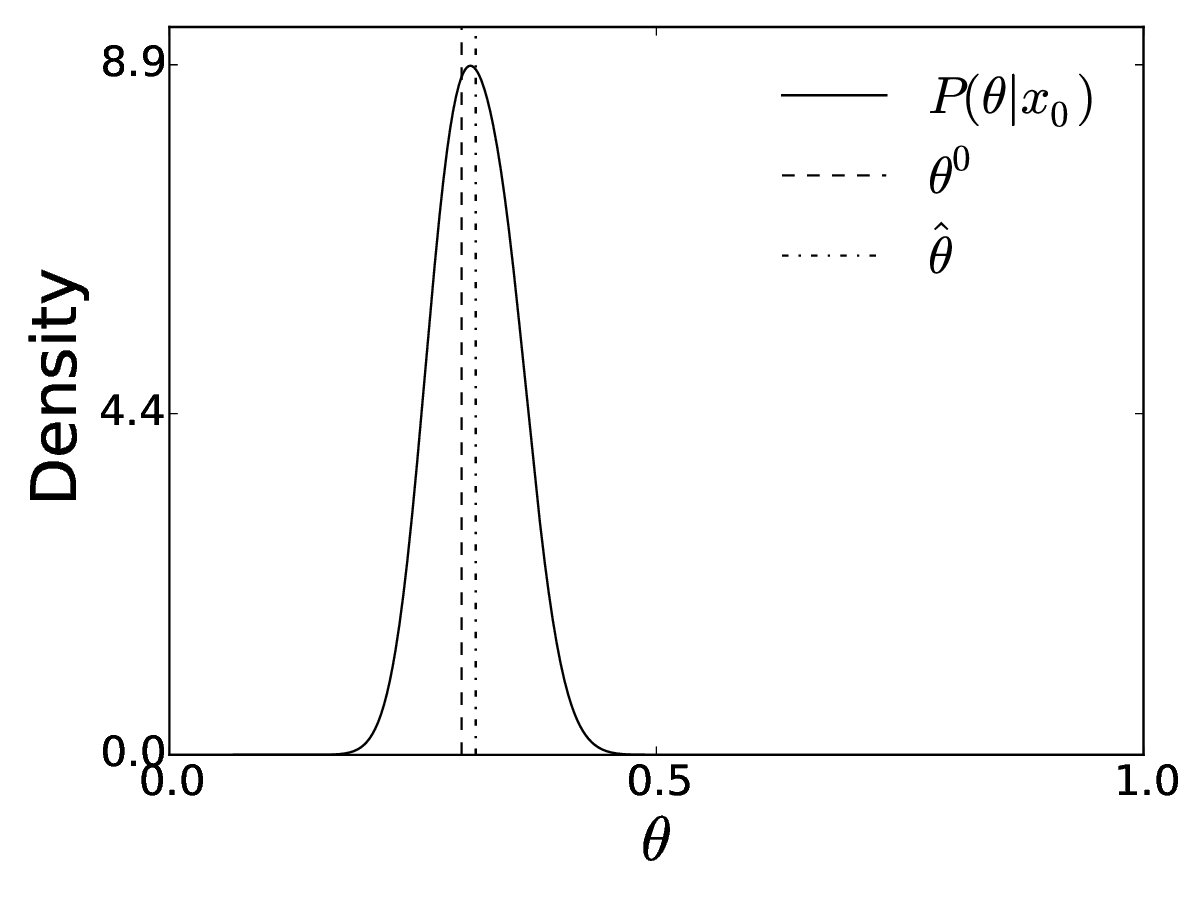}
    \label{fig:ivc_SC_posterior_theta}
    \adjustbox{valign=T}{\subfigure{b.}}
    \adjincludegraphics[valign=T,scale=0.2]{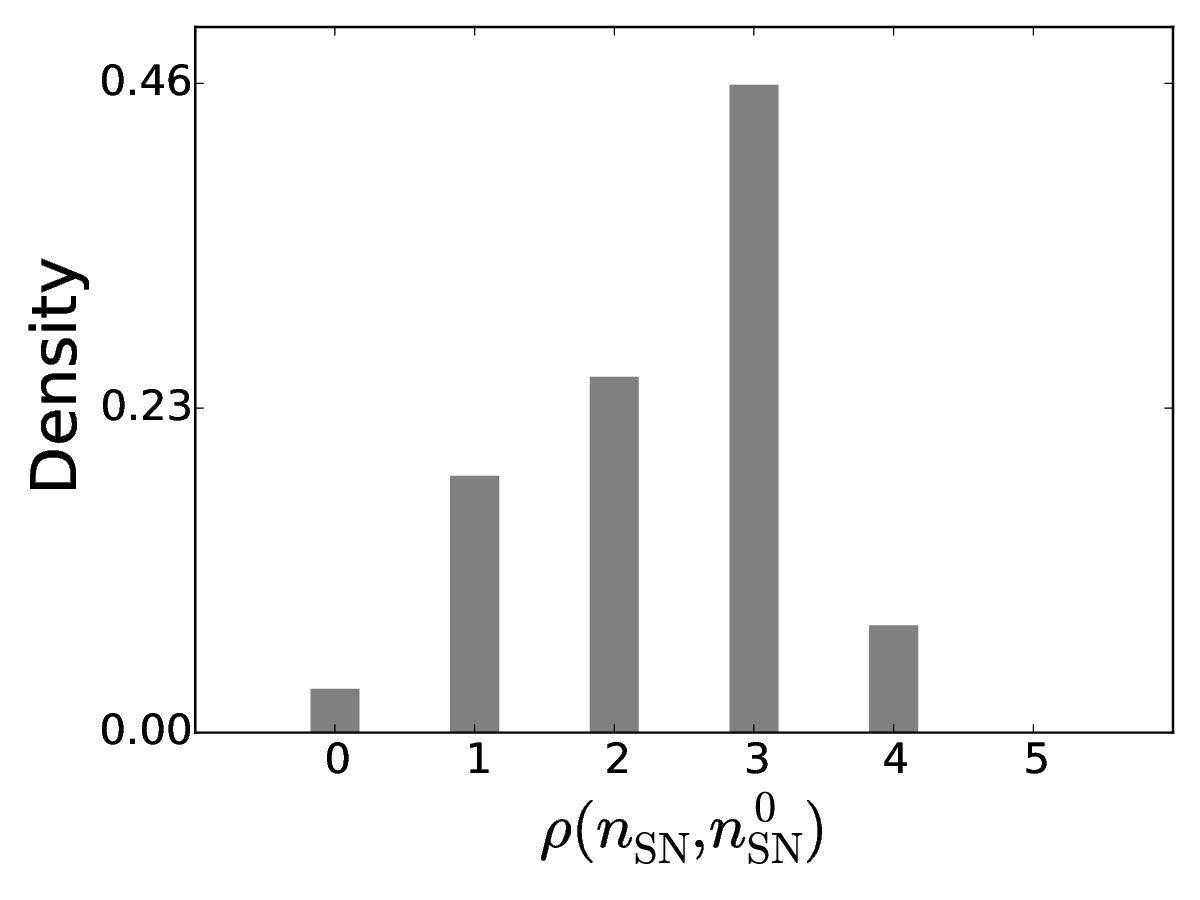}
    \label{fig:ivc_SC_posterior_seednode}
    \adjustbox{valign=T}{\subfigure{c.}}
    \adjincludegraphics[valign=T,scale=0.2]{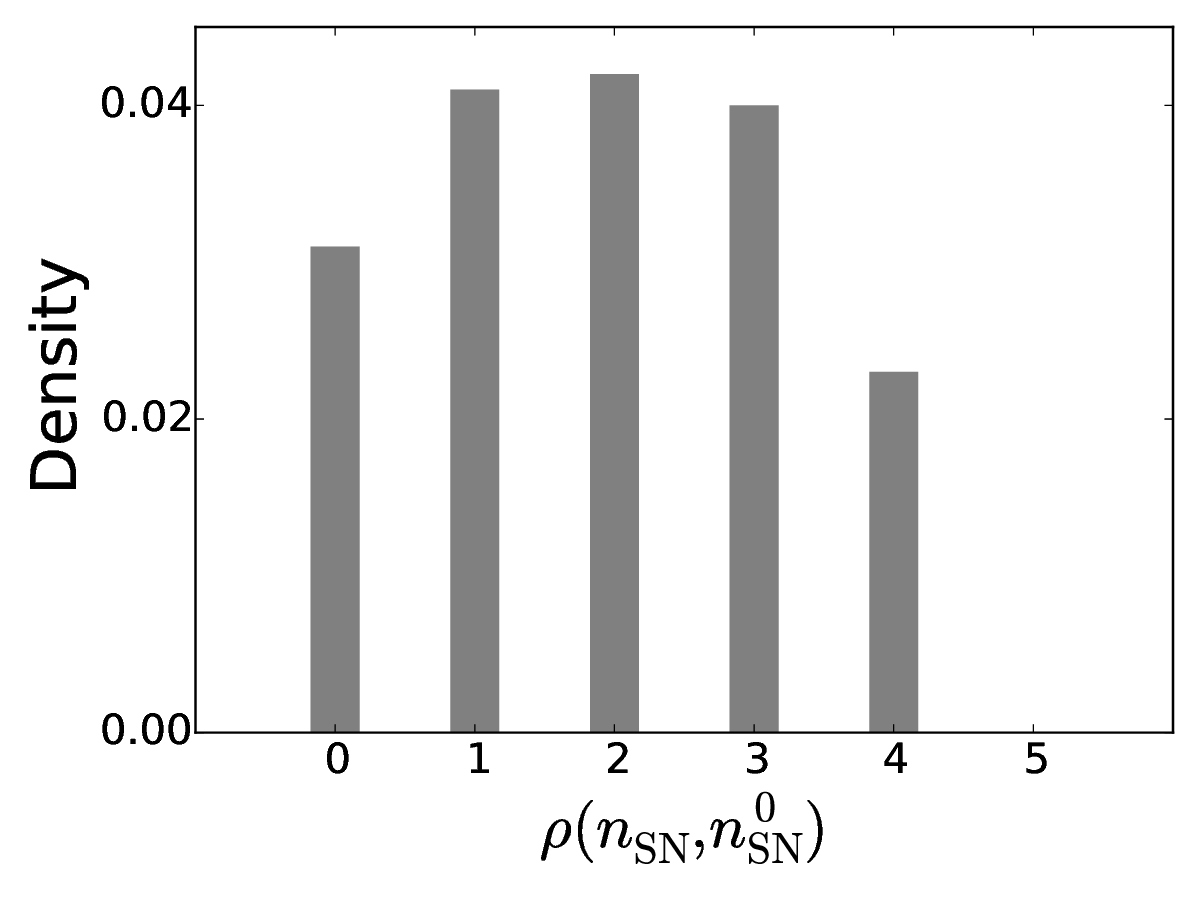}
    \label{fig:ivc_SC_posterior_seednode_max}
\caption{\textbf{Simple contagion model on Indian village contact network.} Panel \textbf{a} shows the density of the inferred marginal posterior distribution and Bayes estimate of $\ratediffusion$, given $\dataObs$, the epidemics on the Indian village contact network. Panel \textbf{b} and  \textbf{c} correspondingly display the marginal posterior distribution and maximum posterior distribution on a single node at different distances from the true seed-node $\seednode^0$. The shortest path length distance between $\seednode^0$ and $\estseednode$ is 1. The diameter of the component, containing $\seednode^0$, of Indian village contact network is 7.}
\label{fig:ivc_SC_posterior_BE}
  \end{figure}

\begin{figure}[htbp]
    \centering
    \mbox{}
    \adjustbox{valign=T}{\subfigure{a.}}
    \adjincludegraphics[valign=T,scale=0.2]{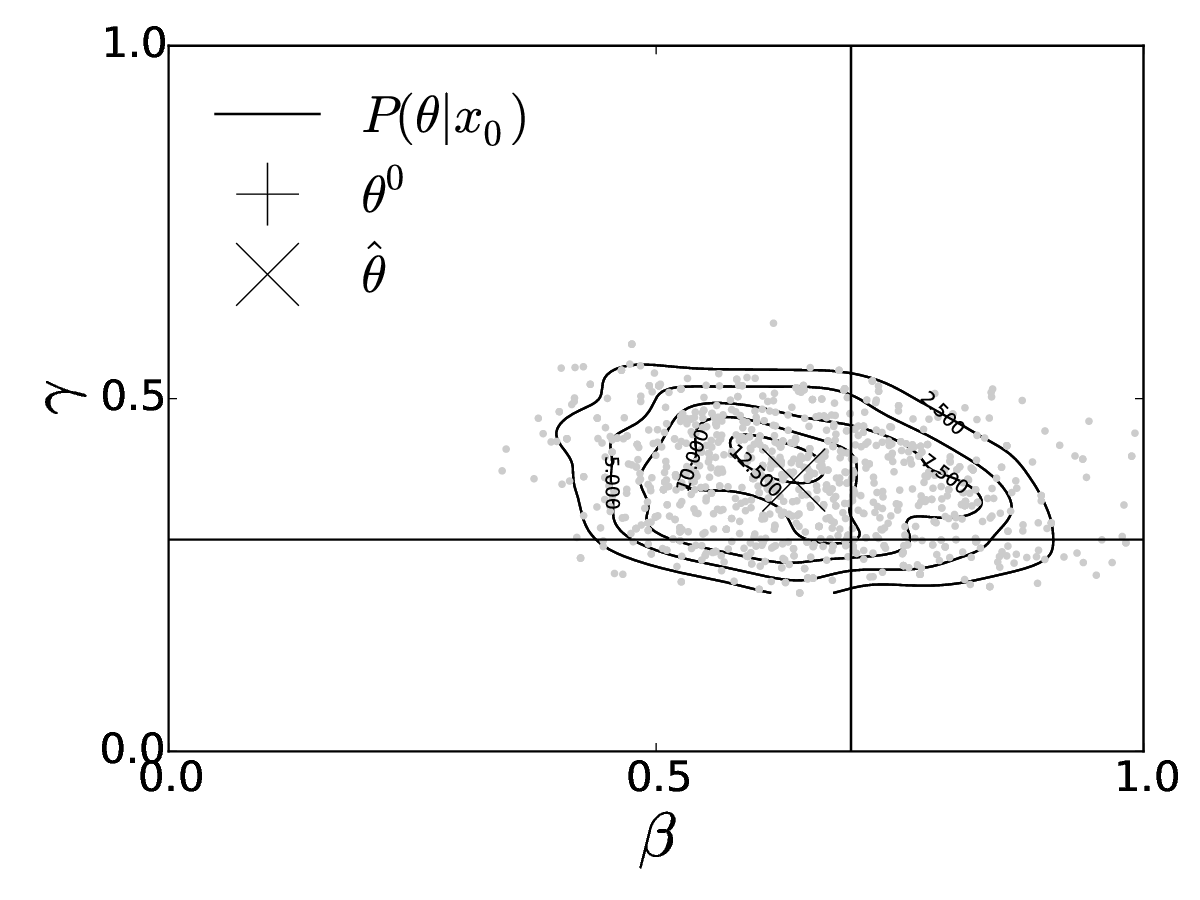}
    \label{fig:fbs_CC_posterior_beta_gamma}
    \adjustbox{valign=T}{\subfigure{b.}}
    \adjincludegraphics[valign=T,scale=0.2]{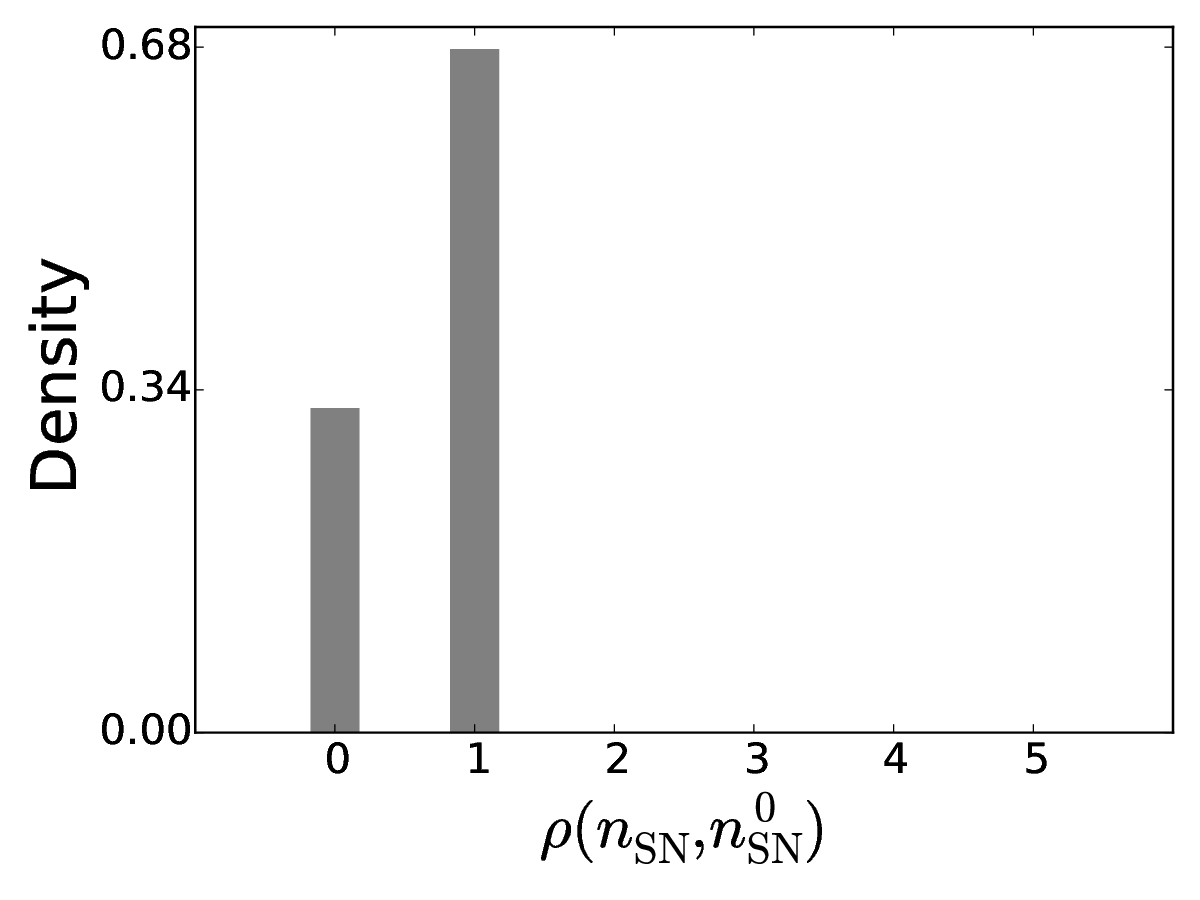}
    \label{fig:fbs_CC_posterior_seednode}
    \adjustbox{valign=T}{\subfigure{c.}}
    \adjincludegraphics[valign=T,scale=0.2]{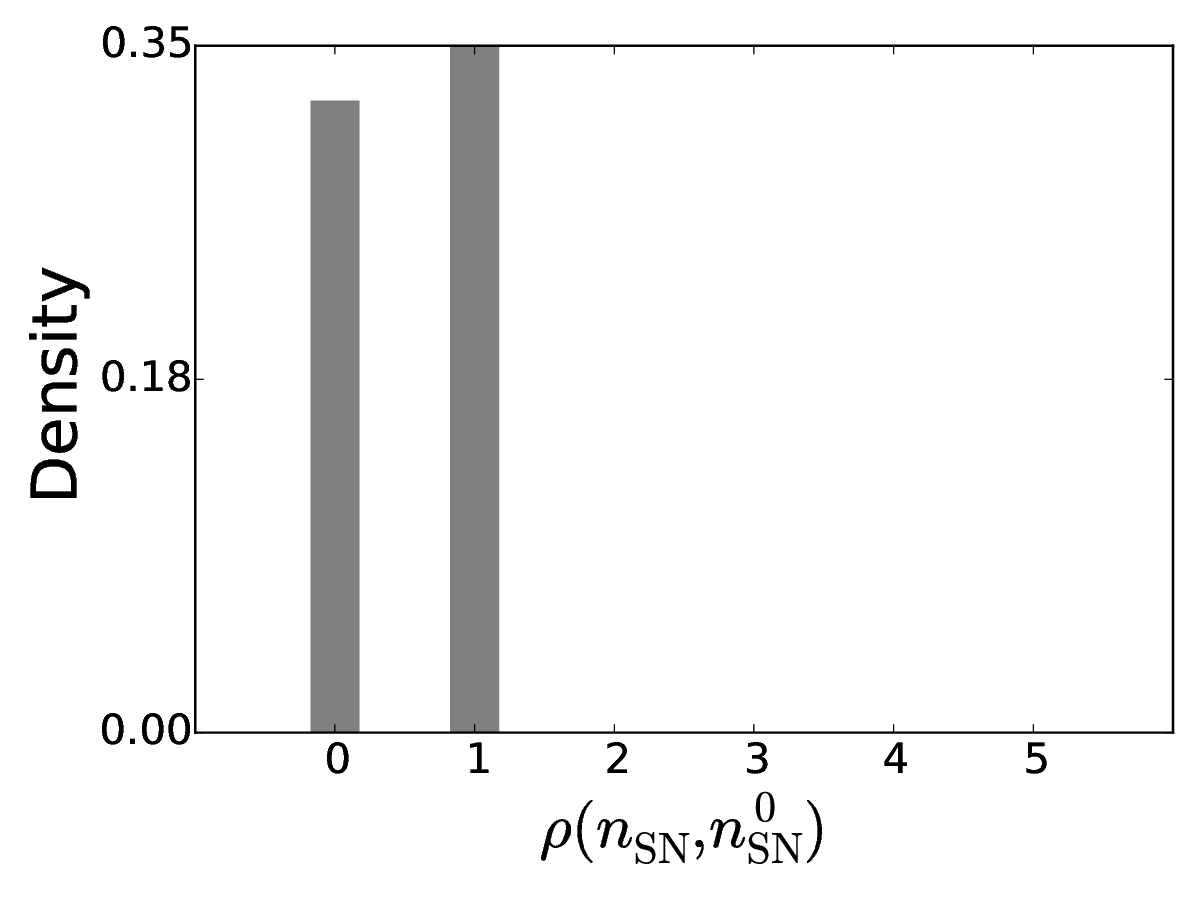}
    \label{fig:fbs_CC_posterior_seednode_max}
\caption{\textbf{Complex contagion model on Facebook Social Network.} Panel \textbf{a} shows the density of the inferred marginal posterior distribution and Bayes estimate of $\parameter$, given $\dataObs$, the epidemics on the Facebook social network. Panel \textbf{b} and \textbf{c} correspondingly display the marginal posterior distribution and maximum posterior distribution on a single node at different distances from the true seed-node $\seednode^0$. The shortest path length distance between $\seednode^0$ and $\estseednode$ is 1. The diameter of the Facebook Social Network is 8.}
\label{fig:fbs_CC_posterior_BE}
  \end{figure}

\begin{figure}[htbp]
    \centering
    \mbox{}
    \adjustbox{valign=T}{\subfigure{a.}}
    \adjincludegraphics[valign=T,scale=0.2]{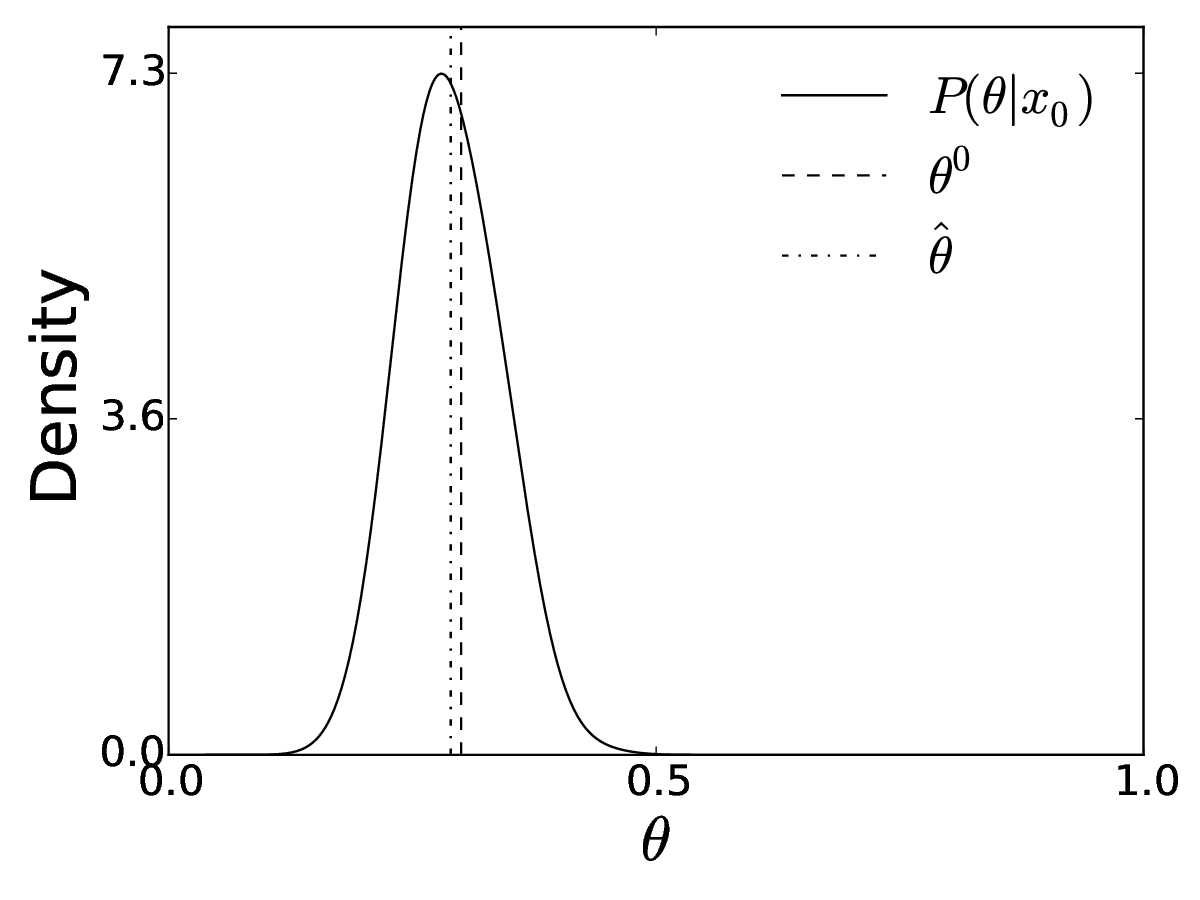}
    \label{fig:fbs_SC_posterior_beta_gamma}
    \adjustbox{valign=T}{\subfigure{b.}}
    \adjincludegraphics[valign=T,scale=0.2]{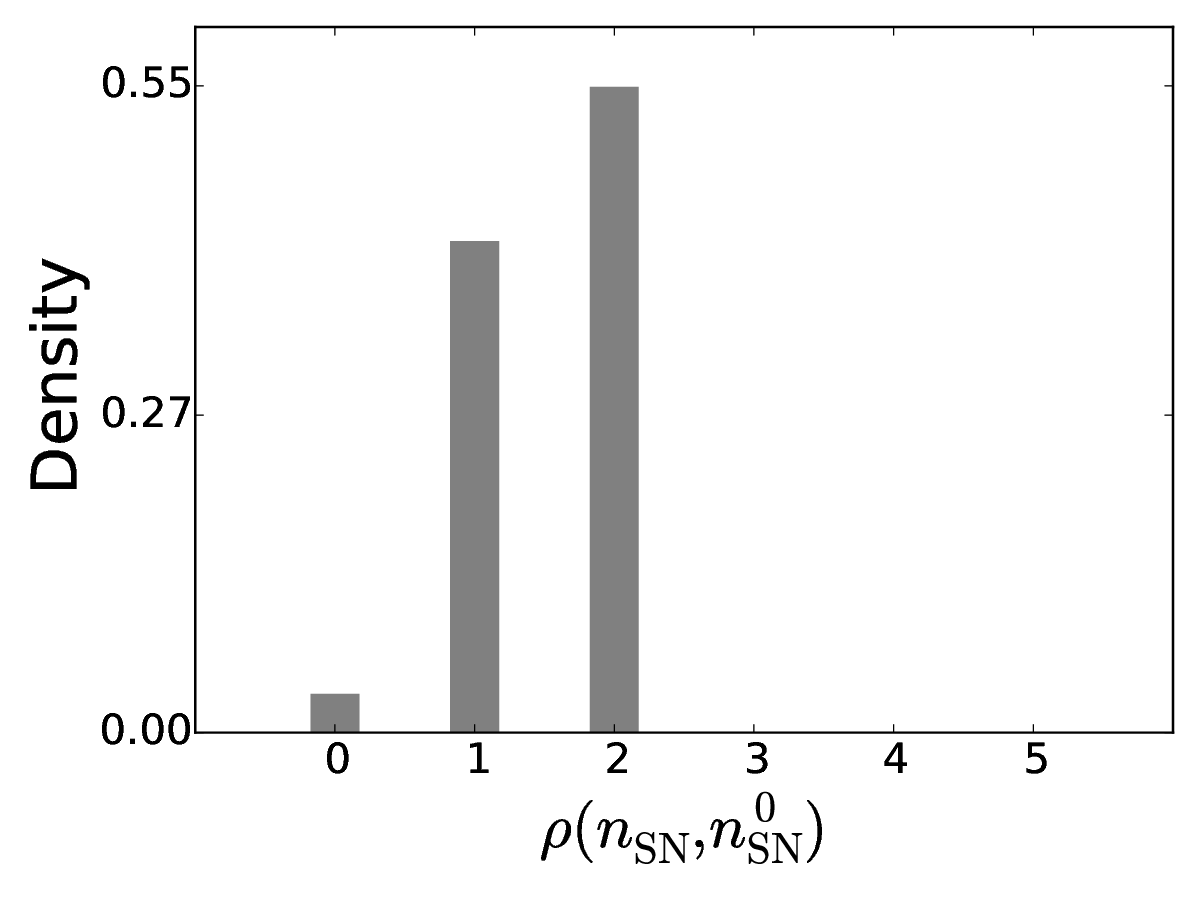}
    \label{fig:fbs_SC_posterior_seednode}
    \adjustbox{valign=T}{\subfigure{c.}}
    \adjincludegraphics[valign=T,scale=0.2]{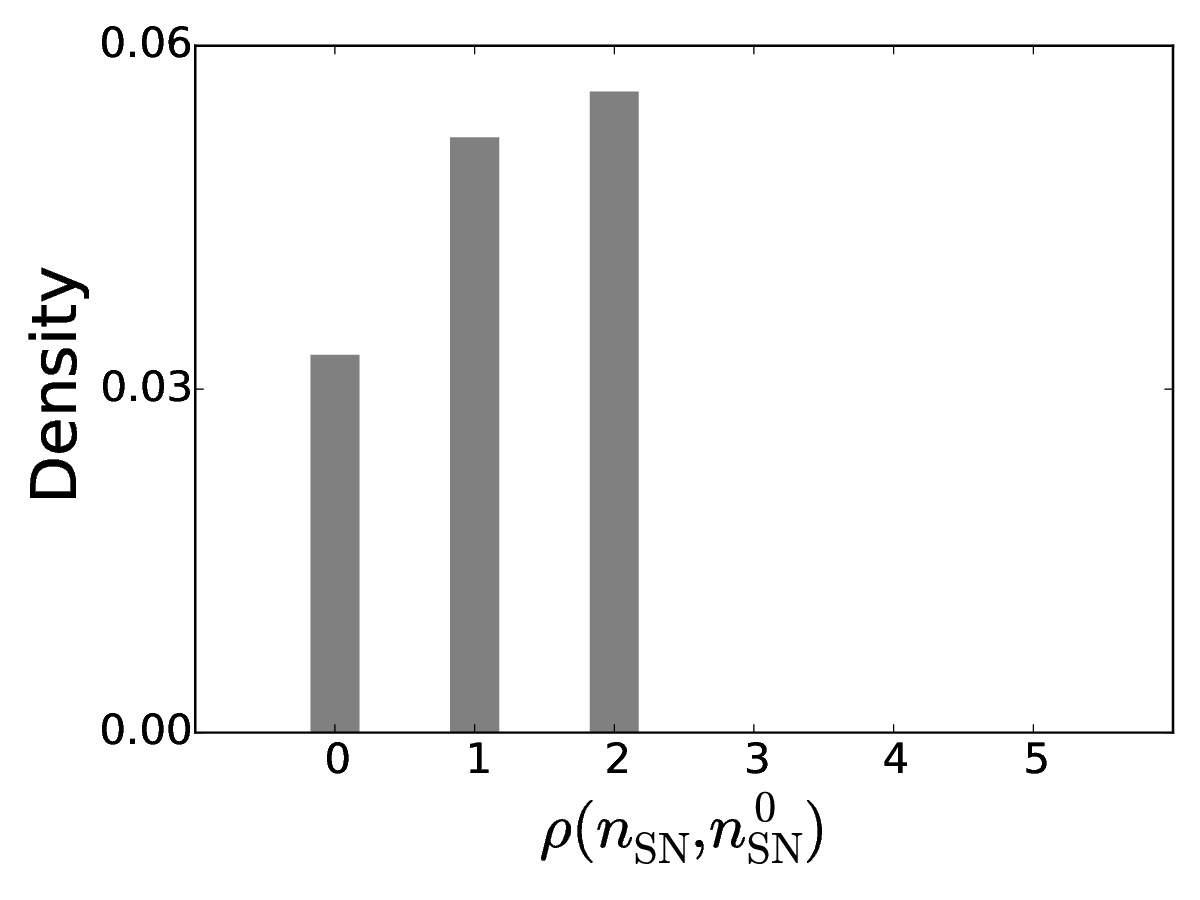}
    \label{fig:fbs_SC_posterior_seednode_max}
\caption{\textbf{Simple contagion model on Facebook Social Network.} Panel \textbf{a} shows the density of the inferred marginal posterior distribution and Bayes estimate of $\ratediffusion$, given $\dataObs$, the epidemics on the Facebook social network. Panel \textbf{b} and \textbf{c} correspondingly display the marginal posterior distribution and maximum posterior distribution on a single node at different distances from the true seed-node $\seednode^0$. The shortest path length distance between $\seednode^0$ and $\estseednode$ is 1. The diameter of the Facebook Social Network is 8.}
\label{fig:fbs_SC_posterior_BE}
  \end{figure}

\begin{figure}[htbp]
    \centering
    \mbox{}%
    \adjustbox{}{\subfigure{a.}}
    \adjincludegraphics[valign=T,scale=0.2]{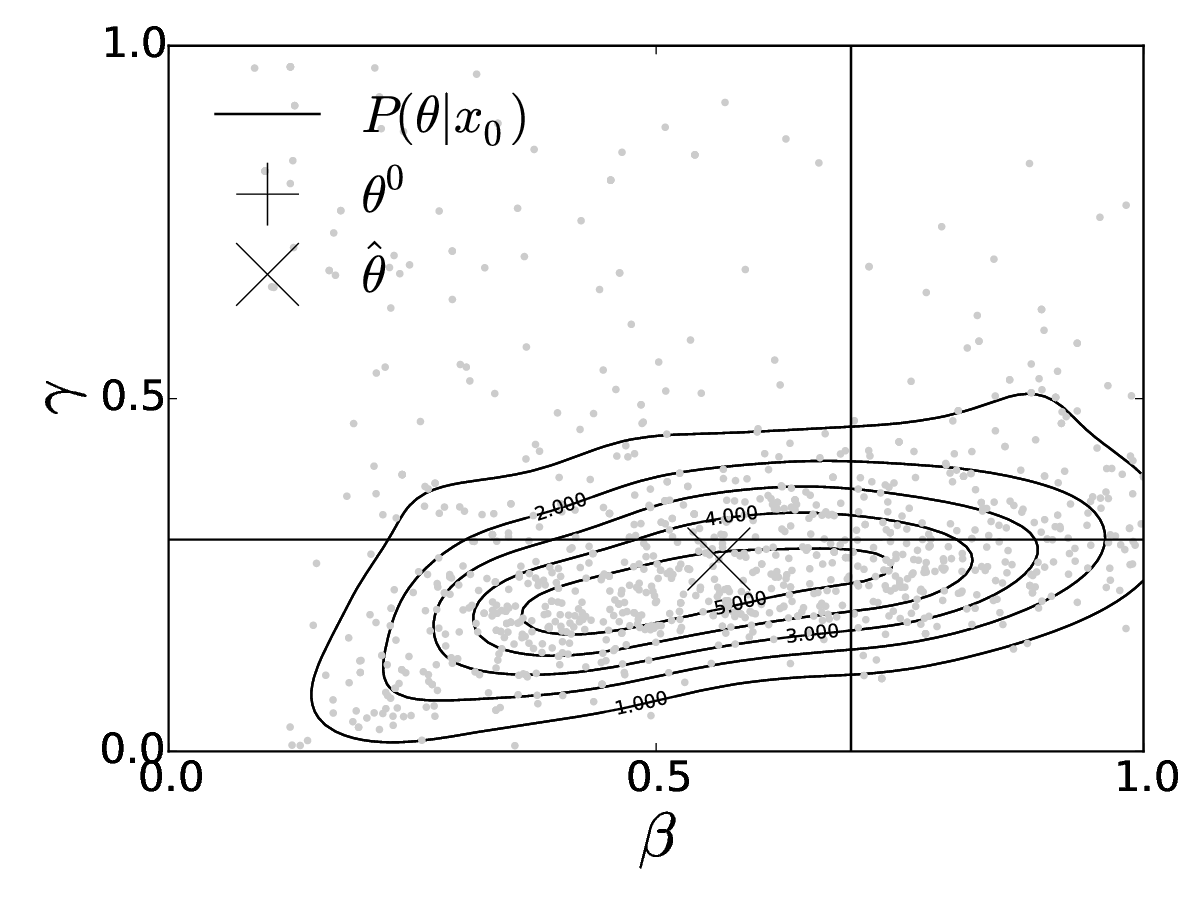}
    \label{fig:ivc_CC_posterior_theta}
    \adjustbox{}{\subfigure{b.}}
    \adjincludegraphics[valign=T,scale=0.2]{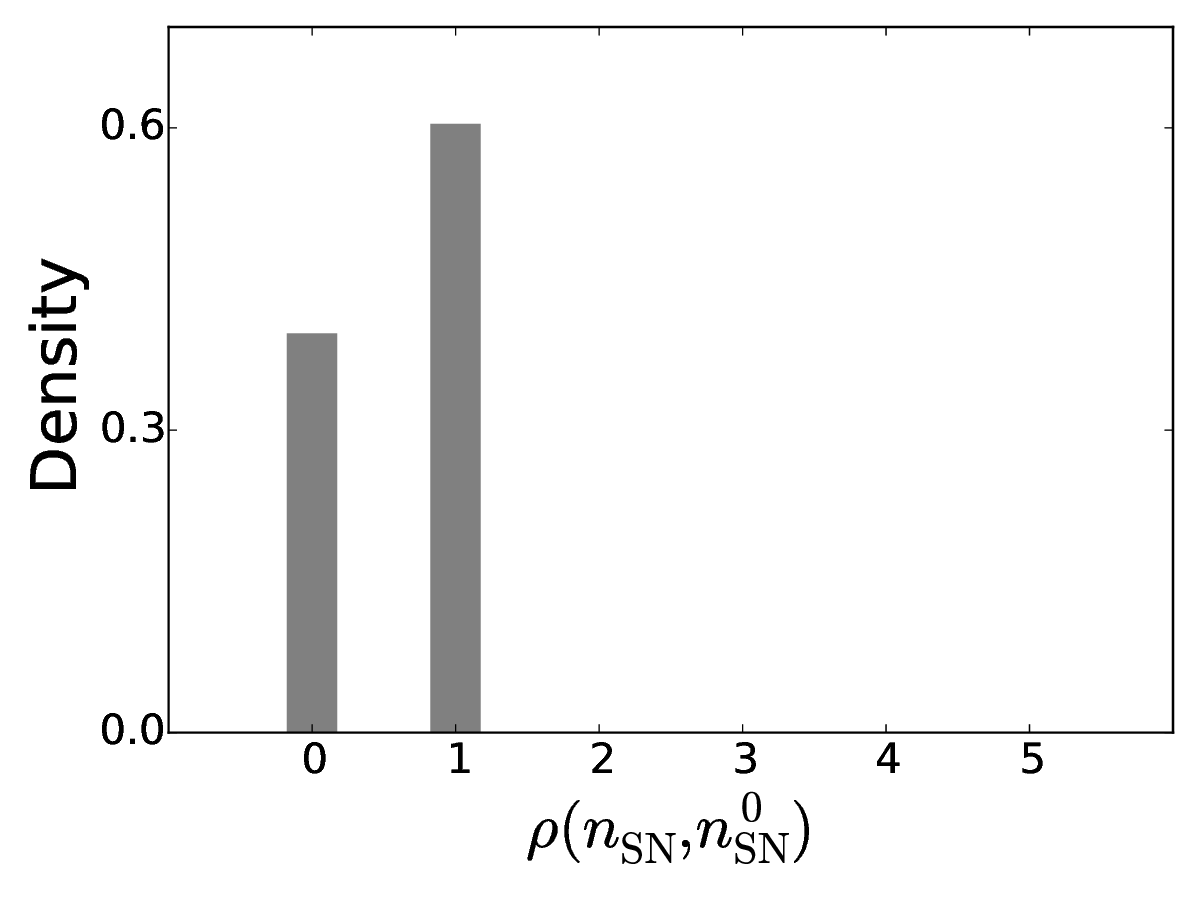}
    \label{fig:ivc_CC_posterior_seednode}
    \adjustbox{}{\subfigure{c.}}
    \adjincludegraphics[valign=T,scale=0.2]{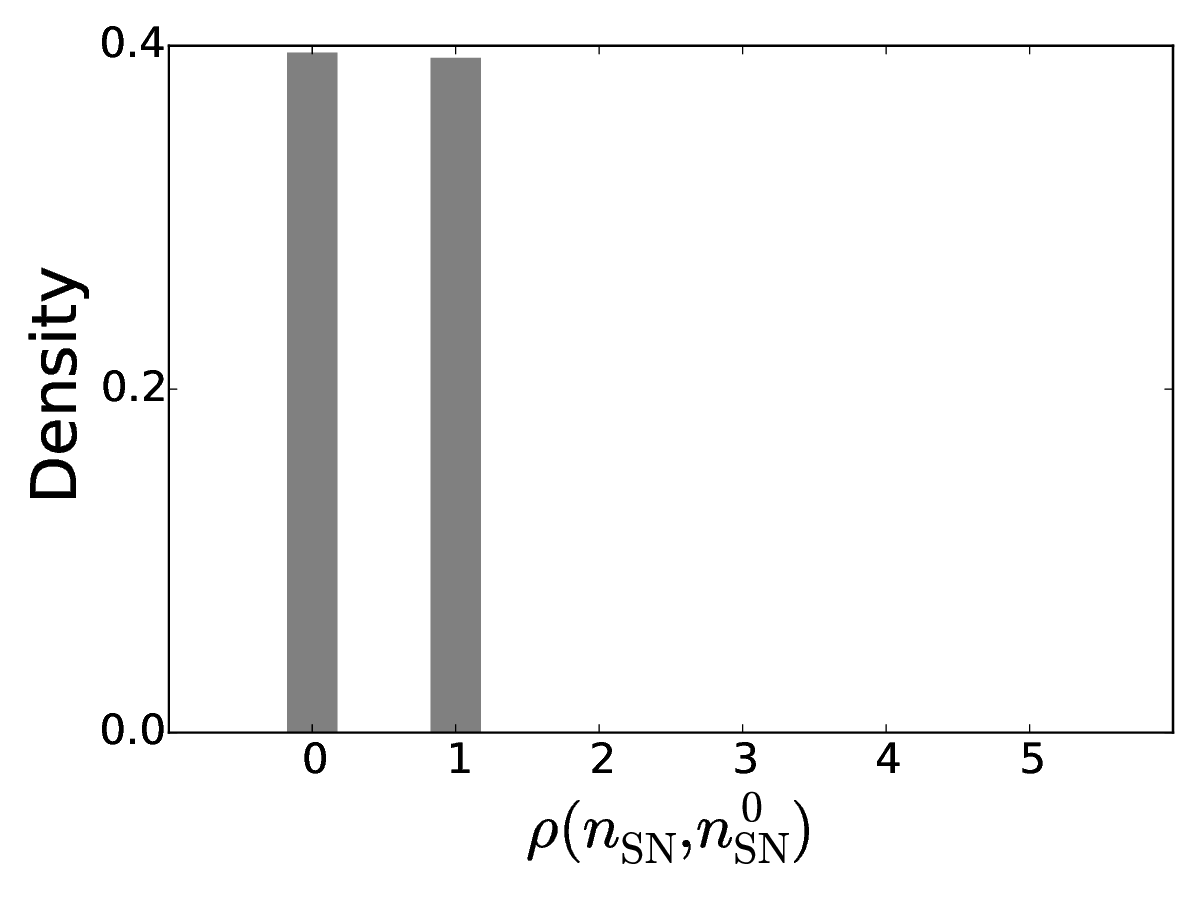}
    \label{fig:ivc_CC_posterior_seednode_max}
\caption{\textbf{Complex contagion model on Indian village contact network.} Panel \textbf{a} shows the density of the inferred marginal posterior distribution and Bayes estimate of $\parameter$, given $\dataObs$, the epidemics on the Indian village contact network. Panel \textbf{b} and  \textbf{c} correspondingly display the marginal posterior distribution and maximum posterior distribution on a single node at different distances from the true seed-node $\seednode^0$. The shortest path length distance between $\seednode^0$ and $\estseednode$ is 0. The diameter of the component, containing $\seednode^0$, of Indian village contact network is 7.}
\label{fig:ivc_CC_posterior_BE}
  \end{figure}

We next turn to inference of the posterior distribution of $\parametertheo \equiv (\parameter, \seednode)$ for epidemics on empirical networks. The marginal posterior distributions and the Bayes estimates of $(\parameter, \seednode)$ are show in Figures~\ref{fig:ivc_SC_posterior_BE}, \ref{fig:fbs_CC_posterior_BE}, \ref{fig:ivc_CC_posterior_BE} \& \ref{fig:fbs_SC_posterior_BE}. 
The inferred posterior distributions for the simple and complex contagion epidemics, both on the Indian village contact network and Facebook social network, is concentrated around the true parameter values. The Bayes estimates are also in a very small neighborhood of the true value, specifically the estimated seed-node $(\estseednode)$ has a shortest path  distance less than or equal to 1 from $\seednode^0$ in all the cases. These two real networks have high average clustering coefficients (Indian village contact network: 0.642, Facebook social network: 0.605), compared to the average clustering
coefficients of the simulated networks considered before (BA network: 0.161, ER network: 0.091). Our results thus demonstrate that the inference scheme can also be successfully applied to highly clustered networks.

\section*{Discussion}
We have introduced a likelihood-free Bayesian inference scheme for studying spreading processes on networks. The framework is 
able to estimate jointly the parameters of the spreading process and the identity of the seed-node, and it is also able to quantify the associated uncertainty. We show that our inferential framework is robust to different network topologies, contagion processes and length of the observation period (number of observations). Both in the synthetic and the empirical applications, the Bayes estimators of the parameters and the seed-nodes are concentrated around their respective true values. 
In general, we find that performing inference on spreading process parameters is considerably easier than performing inference on the seed-nodes, as the variance of the marginal posterior density of spreading process parameters are smaller than the one of the seed-node, while we use uniform prior density for both of them. 
This is intuitive since inference on the infection seed-nodes only makes use of the observations in a single time step (the first one in a sequence of many), whereas inference on spreading process parameters can leverage information present in the entire sequence of observations. 
This asymmetry might also be due to the choice of the summary statistics and the distance measure, which may contain more information on the spreading process parameters.
We also find that inference is generally more difficult on more homogeneous networks, whether model generated (synthetic) or empirical. On synthetic networks, our inference scheme was more accurate for spreading processes propagating on a \ba network with a fat-tailed power-law degree distribution than an \er network with a more concentrated Poisson degree distribution. Similarly, our framework performed better on the empirical Facebook network than the Indian village network, where the latter has a more homogeneous degree distribution due to the use of name generators in network ascertainment, which impose an upper bound on the number of acquaintances that may be nominated  \cite{holland1973structural}, which has been shown to affect the speed and size of epidemics simulated on such networks \cite{harling2016impact}. 
The flexibility of our method allows for many generalizations. For example, one could incorporate nodal covariates in the spreading process to account for different levels of epidemic risk and social influence. 
The most important limitation of our approach is the assumption of observing all infected persons at all time steps, as well as the assumption that the network structure is fully known. Fortunately there is a large literature on Bayesian methods on missing data, and incorporating some of these approaches in our inference scheme is a promising direction for future research in this area. 
Given the generality of the Bayesian inference scheme, the methodology presented here can be extended to situations where we have more than one seed-node \cite{prakash2012spotting}, we have only noisy/partial observations \cite{sundareisan2015hidden} at fixed time points, or when the starting time of the spreading process is not known. Past research on seed-node inference has assumed that the exact infection times and the transmission delays along each edge of the network are known and are jointly distributed following a fixed parametric distribution \cite{pinto2012locating, farajtabar2015back, spinelli2017back}. Though in this manuscript we consider epidemics for which the state of the nodes are observed at some discrete time points and threshold models \cite{GranovetterThreshold_1978, centola2007complex, Fink_AAAI_2016} which may not satisfy the assumption of fixed distribution for transmission delays along all the edges. Our methodology can be also applied to the continuous time processes studied in \cite{pinto2012locating, farajtabar2015back, spinelli2017back}.

\section{Acknowledgment}
The research was supported by Swiss National Science Foundation Grant No. 105218\_163196 (Statistical Inference on Large-Scale Mechanistic Network Models). We thank Dr. Marcel Schoengens, CSCS, ETH Zurich for help regarding HPC and the Swiss National Super Computing Center for providing computing resources.\\\newline
\textbf{Author Contribution:} RD, JPO and AM designed research; RD performed research; RD, JPO and AM wrote the paper.

\bibliographystyle{unsrt}


\end{document}